\documentclass[lettersize,journal]{IEEEtran}
\usepackage{amsmath,amsfonts}
\usepackage{algorithm}
\usepackage{algpseudocode}
\usepackage{array}
\usepackage[caption=false,font=normalsize,labelfont=sf,textfont=sf]{subfig}
\usepackage{textcomp}
\usepackage{stfloats}
\usepackage{url}
\usepackage{verbatim}
\usepackage{graphicx}
\usepackage{cite}
\graphicspath{images/} 
\usepackage{booktabs}
\usepackage{float}
\usepackage{enumitem}
\usepackage{amssymb}
\usepackage{xcolor}
\usepackage{csquotes}
\usepackage{multirow}
\usepackage{booktabs}
\usepackage{makecell}
\usepackage{tikz}
\usetikzlibrary{positioning, arrows.meta}

\begin{document}

\title{Low-Complexity Neural Pursuit for Line Spectral Estimation}
\author{
Chenchen Liu,~\IEEEmembership{Graduate Student Member,~IEEE}, 
Wenjun Jiang,~\IEEEmembership{Graduate Student Member,~IEEE}, 
and Xiaojun Yuan,~\IEEEmembership{Fellow,~IEEE}
\vspace{-5pt}
\thanks{A preliminary conference version of this work appeared in \cite{11270737}. This manuscript substantially extends the conference version in methodology, simulation evaluation, and benchmark comparison.}
\thanks{\textcopyright~2026 IEEE. Personal use of this material is permitted. Permission from IEEE must be obtained for all other uses, in any current or future media, including reprinting/republishing this material for advertising or promotional purposes, creating new collective works, for resale or redistribution to servers or lists, or reuse of any copyrighted component of this work in other works. Published version: C. Liu, W. Jiang, and X. Yuan, ``Low-Complexity Neural Pursuit for Line Spectral Estimation,'' \emph{IEEE Transactions on Signal Processing}, vol.~74, pp.~3317--3332, 2026, doi: 10.1109/TSP.2026.3715966.}
}




\maketitle

\begin{abstract}
    In this paper, we propose a high-precision yet low-complexity framework, termed Neural Pursuit for Line Spectral Estimation (NeuPLSE), for off-grid line spectral estimation from a single snapshot. 
    The proposed framework systematically integrates model-driven signal structures with data-driven adaptability, enabling continuous-domain spectral representation together with adaptive parameter learning. 
    Within this framework, model order detection and spectral parameter estimation are jointly achieved through residual analysis, with coordinated use of global and generalized component-wise residual information to ensure robust and accurate estimation performance.
    Moreover, NeuPLSE supports a unified extension from one-dimensional (1D) to multi-dimensional (MD) scenarios while maintaining low computational complexity.
    Extensive simulations in representative integrated sensing and communication (ISAC) scenarios demonstrate that NeuPLSE achieves accurate model order detection and high-precision spectral estimation. In sparse-path settings, it approaches the Cramér--Rao lower bound, while in dense-path settings, it attains competitive reconstruction performance with significantly lower computational complexity than state-of-the-art methods.
\end{abstract}

\begin{IEEEkeywords}
	line spectral estimation, residual analysis, model order detection, low-complexity estimation
\end{IEEEkeywords}

\section{Introduction}
Line spectral estimation (LSE) is a fundamental problem in statistical signal processing \cite{stoica2005spectral}, with wide-ranging applications in radar and sonar parameter estimation \cite{cui2025linearly,teng2022bayesian,10815057}, wireless channel estimation \cite{jiang2025interference}, medical imaging \cite{6930780}, and speech signal processing \cite{jang2011efficient}. 
Integrated sensing and communication (ISAC) \cite{wei2023integrated} serves as a representative application, further emphasizing the role of LSE in modern wireless systems. 
With the increasing system scale and the demand for real-time processing in such scenarios, developing LSE methods that simultaneously achieve high estimation accuracy and low computational complexity has become crucial \cite{saad2019vision,luong2025advanced}.

A variety of methods have been developed for LSE over the past decades. 
Early approaches based on the discrete Fourier transform (DFT) identify spectral components on a discrete frequency grid \cite{stoica2005spectral}, resulting in inherent resolution limitations. 
Parametric methods have been extensively developed to improve estimation accuracy.
The maximum likelihood (ML) estimator \cite{ottersten1992analysis} achieves asymptotically optimal performance when the model order is known, but requires solving a computationally intensive nonconvex optimization problem. 
Subspace-based methods, such as MUSIC \cite{schmidt1986multiple} and ESPRIT \cite{roy1989esprit}, exploit the covariance eigenstructure to achieve super-resolution estimation without grid discretization. 
However, they rely on multiple-snapshot observations for accurate covariance estimation, and their performance degrades in low signal-to-noise ratio (SNR) or limited-snapshot regimes.

Sparse modeling offers an alternative framework for LSE by exploiting spectral sparsity in the frequency domain. 
By discretizing the continuous frequency space onto a predefined grid, the problem can be formulated within the compressed sensing (CS) framework \cite{hu2012compressed,5437428,han2025dmra}. 
This formulation allows the use of sparse recovery algorithms such as orthogonal matching pursuit (OMP) \cite{cai2011orthogonal}, the least absolute shrinkage and selection operator (LASSO) \cite{tibshirani1996regression}, and alternating direction method of multipliers (ADMM) \cite{neal2011distributed} to estimate spectral components. 
However, grid discretization leads to off-grid errors when the true frequencies do not coincide with the predefined grid points, resulting in performance degradation. 
While increasing the grid density can partially alleviate this issue, it incurs significantly higher computational complexity.

The mismatch induced by grid discretization has motivated the development of approaches operating directly in the continuous parameter domain.
Atomic norm minimization (ANM) \cite{bhaskar2013atomic, tang2014near} formulates LSE as a convex optimization problem over a continuous dictionary, thereby eliminating discretization errors, at the cost of solving computationally intensive semidefinite programs (SDP).
Bayesian approaches, such as variational line spectral estimation (VALSE) \cite{badiu2017variational}, achieve high estimation accuracy together with automatic model order selection, but often incur substantial computational complexity. 
To reduce this complexity, Superfast LSE \cite{hansen2018superfast} introduces an accelerated implementation. 
However, this acceleration relies on structural properties specific to one-dimensional (1D) settings, and its extension to multi-dimensional (MD) scenarios remains nontrivial due to the more complex Block-Toeplitz with Toeplitz Blocks (BTTB) covariance structure.

In recent years, deep learning (DL)-based approaches have been extensively investigated for LSE.
Existing DL-based methods can be broadly categorized into purely data-driven and hybrid-driven paradigms.
Purely data-driven approaches aim to learn a direct mapping from observations to spectral parameters.
These methods typically formulate LSE either as a regression problem for continuous frequency estimation \cite{adavanne2018direction, guo2020doa} or as a classification task over discretized frequency grids \cite{chakrabarty2019multi, tian2022vehicle}.
To handle unknown model order, several adaptive strategies have been explored, including subregion-based decomposition with parallel networks \cite{elbir2020deepmusic}, model-order-specific training \cite{yu2023deep, wang2024effective}, and multi-label classification schemes \cite{papageorgiou2021deep, zheng2024deep, naoumi2024complex}.
Despite their flexibility, these approaches generally suffer from limited interpretability and often rely on discretized grids, which may lead to model mismatch.

Hybrid-driven approaches integrate model-based signal processing principles with neural networks and can be further divided into two classes. 
The first class includes heuristic combinations of neural networks and classical signal processing modules, such as methods built upon MUSIC \cite{jiang2019deep, 10266765, xu2025deep}.
Although such methods incorporate useful domain knowledge, they are generally not derived from a unified estimation or optimization framework, which limits the depth of integration between model-based structure and data-driven learning. 
The second class is deep unfolding, which offers a more principled way to integrate model-based and data-driven designs by transforming iterative algorithms into trainable network architectures. 
Existing deep unfolding methods for LSE can be further divided into two categories: on-grid and off-grid approaches. 
On-grid methods are typically derived from sparse recovery algorithms under discretized dictionaries, with representative examples including LISTA-Net \cite{niu2025direction}, IRLS-Net \cite{jin2025sparse}, and ADMM-Net \cite{wang2024single}.
However, they still suffer from off-grid mismatch due to grid discretization.  
Off-grid approaches aim to estimate spectral parameters directly in the continuous domain without predefined grids, but related studies are mainly based on ANM \cite{raza2025deep, yang2025lanm}. 
To obtain computationally tractable implementations, these methods usually rely on approximate reformulations rather than directly unfolding the original ANM, which reduces computational complexity but generally incurs estimation performance loss.  
Moreover, ANM-based learning methods still suffer from the scalability issue in MD settings.

Despite the recent progress, existing approaches remain limited in simultaneously achieving high estimation accuracy and low computational complexity for off-grid LSE.
The main difficulty is that, due to the superposition of multiple line spectral components, the LSE problem is highly non-convex and thus susceptible to local optima.
To attain high estimation accuracy, existing high-accuracy methods, such as VALSE, rely on precise initialization together with computationally intensive iterative refinement, thereby incurring high complexity.
As such, it is highly challenging to design a method that can simultaneously achieve high estimation accuracy and low computational complexity for off-grid LSE.

To address the above challenges, we propose a novel hybrid framework, termed \textbf{Neural Pursuit for Line Spectral Estimation (NeuPLSE)}, which systematically integrates model-driven signal structures with data-driven adaptability for off-grid LSE.
The proposed framework consists of an initial coarse estimation step followed by a subsequent iterative refinement process.
By effectively exploiting the underlying parametric signal model in continuous-domain estimation, NeuPLSE incorporates residual analysis into the refinement process to enhance estimation performance.
Further, adaptive model order detection through component selection, regeneration, and fusion alleviates the sensitivity of the algorithm to initialization.
Moreover, the data-driven part introduces learnable parameters for adaptive adjustment of key algorithmic parameters, further improving the robustness of the proposed algorithm.
Through this design, NeuPLSE provides a high-accuracy yet low-complexity framework for off-grid LSE.
Compared with the existing off-grid methods (such as VALSE), the proposed NeuPLSE exhibits much stronger learning capacity in adapting to dynamic model parameters. As a result, NeuPLSE only needs to use a simple DFT operation for initialization and is less prone to getting stuck at local optima.

\textbf{The main contributions of this work are summarized as follows:}

\begin{itemize}
\item \textbf{A hybrid model-driven and data-driven framework for off-grid line spectral estimation:} 
We establish a structured framework for off-grid LSE under single-snapshot observation, where model-driven signal structures and data-driven adaptability are systematically integrated. 
Specifically, the model-driven part exploits the underlying parametric signal model for continuous-domain representation, while the data-driven part introduces learnable parameterization to enable adaptive adjustment of key algorithmic parameters. 
Within this framework, different operations are coherently organized to fulfill distinct roles in the estimation process, including coarse initialization, parameter refinement, and model order detection.

\item \textbf{Joint model order detection and spectral parameter refinement via residual analysis:}
The NeuPLSE framework jointly performs model order detection and spectral parameter refinement through systematic residual analysis. 
Global residual captures the overall reconstruction mismatch and guides model order updates via component regeneration, selection, and fusion. 
Generalized component-wise residual characterizes local mismatches associated with individual components, enabling the refinement of frequencies and complex amplitudes.

\item \textbf{A unified extension from one-dimensional to multi-dimensional scenarios with low complexity:}
The NeuPLSE framework admits a unified extension from 1D to MD scenarios while preserving the same core estimation mechanism.
Only the signal representation and parameter updates are generalized to higher-dimensional settings, with low computational complexity maintained.

\item \textbf{Comprehensive evaluation in representative scenarios:}  
The proposed framework is comprehensively evaluated through extensive numerical studies, where ISAC is adopted as a representative application scenario. 
NeuPLSE achieves accurate model order detection and spectral estimation while maintaining favorable computational efficiency. 
In sparse-path settings, it approaches the Cramér--Rao lower bound (CRLB), while in dense-path scenarios it attains competitive reconstruction accuracy relative to VALSE. 
In addition, ablation studies of the core mechanisms, sensitivity analyses of key hyperparameter settings, and generalization evaluations under mismatched distributions of path numbers are conducted to comprehensively evaluate NeuPLSE.

\end{itemize}

\textit{Organization:}  
The system model and problem formulation of LSE are introduced in Section~\ref{Sec_SM}, followed by the NeuPLSE framework in Section~\ref{NeuPLSE_framework}. 
The 1D implementation of NeuPLSE is presented in Section~\ref{Sec_1DNeuPLSE}, while its MD implementation is developed in Section~\ref{Sec_MDNeuPLSE}. 
Numerical results are provided in Section~\ref{sec:simulation}. 
Finally, Section~\ref{sec:conclusion} concludes the paper and outlines future research directions.

\textit{Notation:} 
Bold lowercase and uppercase letters (e.g., $\mathbf{x}$ and $\mathbf{X}$) denote vectors and matrices, respectively. 
The $\ell_2$ norm of a vector $\mathbf{x}$ is denoted by $\|\mathbf{x}\|_2$, and $|\cdot|$ denotes the absolute value of a scalar. 
The operators $(\cdot)^{T}$, $(\cdot)^{*}$, and $(\cdot)^{H}$ denote the transpose, conjugate, and conjugate transpose, respectively. 
The set of $M \times N$ complex matrices is denoted by $\mathbb{C}^{M \times N}$. 
The real and imaginary parts of a complex number are denoted by $\operatorname{Re}(\cdot)$ and $\operatorname{Im}(\cdot)$, respectively. 
$\mathrm{Bernoulli}(\cdot)$ denotes the Bernoulli distribution, and $\mathcal{CN}(\cdot;\mu,\Sigma)$ denotes the circularly symmetric complex Gaussian distribution with mean $\mu$ and covariance $\Sigma$. 
The expectation operator is denoted by $\mathbb{E}[\cdot]$, and $\mathcal{O}(\cdot)$ denotes the standard big-$\mathcal{O}$ complexity.

\section{System Model and Problem Formulation} \label{Sec_SM}

In this section, the system models for both 1D-LSE and MD-LSE are introduced. 
The corresponding estimation problem and performance metrics are subsequently defined.

\subsection{System Model for 1D-LSE}

In the 1D-LSE problem, the noiseless signal $\mathbf{x} \in \mathbb{C}^{N}$ is given by
\begin{equation} \label{1D_channel}
    \mathbf{x} = \sum_{l=1}^{L} \alpha_{l} \mathbf{a}(\omega_{l}),
\end{equation}
where $L$ denotes the model order, i.e., the number of complex exponential components. 
The parameters $\alpha_l \in \mathbb{C}$ and $\omega_l \in [0, 2\pi)$ represent the complex amplitude and frequency of the $l$-th component, respectively. 
The normalized steering vector associated with $\omega_l$ is defined as
\begin{equation} \label{steering_vector}
    \mathbf{a}(\omega_{l}) = \frac{1}{\sqrt{N}} \left[1, e^{-i\omega_{l}}, \dots, e^{-i(N-1)\omega_{l}}\right]^T.
\end{equation}
The observed signal is given by
\begin{equation} \label{1D_measurements}
    \mathbf{y} = \mathbf{x} + \mathbf{n},
\end{equation}
where $\mathbf{n} \sim \mathcal{CN}(\mathbf{0}, \sigma^2 \mathbf{I})$ denotes additive white Gaussian noise (AWGN).

In ISAC systems, the 1D-LSE problem corresponds to 1D sensing or channel estimation tasks, where each spectral component represents a channel subpath. 
Specifically, the model order $L$, complex amplitude $\alpha_l$, and frequency $\omega_l$ correspond to the number of subpaths, complex gain, and channel parameters such as time of arrival (ToA), Doppler shift (DS), azimuth angle of arrival (AoA), and elevation angle of arrival (EoA), respectively \cite{tse2005fundamentals}. 
For brevity, these parameters are referred to as ToA, DS, AoA, and EoA.

\subsection{System Model for MD-LSE}

In the MD-LSE problem, the noiseless signal is modeled as a superposition of $L$ complex exponential components across $D$ dimensions. 
Let the sample sizes along each dimension be $N_1, \dots, N_D$, with total sample size $N = \prod_{d=1}^{D} N_d$. 
The vectorized noiseless signal $\mathbf{x} \in \mathbb{C}^{N}$ is given by
\begin{equation} \label{MD_channel}
    \mathbf{x} = \sum_{l=1}^{L} \alpha_{l} \mathbf{a}(\boldsymbol{\omega}_{l}),
\end{equation}
where $\alpha_l \in \mathbb{C}$ is the complex amplitude of the $l$-th component, and $\boldsymbol{\omega}_l = [\omega_l^{1}, \dots, \omega_l^{D}]^T$ denotes its frequency vector with $\omega_l^{d} \in [0, 2\pi)$ along the $d$-th dimension. 
The corresponding MD steering vector is defined via the Kronecker product of 1D steering vectors:
\begin{equation} \label{MD_steering_vector}
    \mathbf{a}(\boldsymbol{\omega}_{l}) = \frac{1}{\sqrt{N}} \left( \mathbf{a}_1(\omega_l^{1}) \otimes \cdots \otimes \mathbf{a}_D(\omega_l^{D}) \right),
\end{equation}
with
\begin{equation} \label{MD_1D_steering}
    \mathbf{a}_d(\omega_l^{d}) = \left[1, e^{-i\omega_l^{d}}, \dots, e^{-i(N_d - 1)\omega_l^{d}}\right]^T.
\end{equation}
The observed signal is given by
\begin{equation} \label{MD_measurements}
    \mathbf{y} = \mathbf{x} + \mathbf{n}, \quad \mathbf{n} \sim \mathcal{CN}(\mathbf{0},\sigma^2 \mathbf{I}).
\end{equation}
The set $\{\boldsymbol{\omega}_l\}_{l=1}^{L}$ is referred to as the MD frequencies, which jointly characterize the spectral components.

In ISAC systems, the MD-LSE problem corresponds to MD sensing or channel estimation tasks, where each spectral component represents a channel subpath. 
The model order $L$ and complex amplitude $\alpha_l$ correspond to the number of subpaths and the complex gain, respectively. 
Each frequency component $\omega_l^{d}$ is associated with a physical parameter such as ToA, DS, AoA, or EoA.

\subsection{Problem Formulation} \label{Problem Formulation}

Given the system models in \eqref{1D_channel}--\eqref{1D_measurements} and \eqref{MD_channel}--\eqref{MD_measurements}, the LSE problem consists in jointly estimating the unknown model order $L$ and the associated spectral parameters from the noisy observation $\mathbf{y}$. 
For 1D-LSE, the parameters to be estimated are $\{(\omega_l,\alpha_l)\}_{l=1}^{L}$, where each component is characterized by a scalar frequency $\omega_l$. 
For MD-LSE, the parameters are $\{(\boldsymbol{\omega}_l,\alpha_l)\}_{l=1}^{L}$, where each component is characterized by a frequency vector $\boldsymbol{\omega}_l$. 
Let $\hat{L}$ denote the estimated model order, with corresponding estimates $\{(\hat{\omega}_l,\hat{\alpha}_l)\}_{l=1}^{\hat{L}}$ for 1D-LSE and $\{(\hat{\boldsymbol{\omega}}_l,\hat{\alpha}_l)\}_{l=1}^{\hat{L}}$ for MD-LSE.

A standard metric for frequency estimation accuracy is the root mean square error (RMSE), defined under the assumption of correct model order detection, i.e., $\hat{L}=L$. 
For the 1D case,
\begin{equation} \label{RMSE_1D}
    \operatorname{RMSE}(\boldsymbol{\omega}) =
    \sqrt{\mathbb{E}\left[\frac{1}{L}\sum_{k=1}^{L}\min_{\pi\in\Pi}
    \left|\omega_k-\hat{\omega}_{\pi(k)}\right|^2\right]},
\end{equation}
where $\boldsymbol{\omega}=[\omega_1,\dots,\omega_L]^T$ and $\hat{\boldsymbol{\omega}}=[\hat{\omega}_1,\dots,\hat{\omega}_{\hat{L}}]^T$ denote the true and estimated frequency vectors, respectively. The minimization over the permutation set $\Pi$ accounts for label ambiguity.

For the MD case,
\begin{equation} \label{RMSE_MD}
    \operatorname{RMSE}(\boldsymbol{\Omega}) =
    \sqrt{\mathbb{E}\left[\frac{1}{LD}\sum_{l=1}^{L}\min_{\pi\in\Pi}
    \left\|\boldsymbol{\omega}_l-\hat{\boldsymbol{\omega}}_{\pi(l)}\right\|_2^2\right]},
\end{equation}
where $\boldsymbol{\Omega} = [\boldsymbol{\omega}_1,\dots,\boldsymbol{\omega}_L]$ and
$\hat{\boldsymbol{\Omega}} = [\hat{\boldsymbol{\omega}}_1,\dots,\hat{\boldsymbol{\omega}}_{\hat{L}}]$
denote the true and estimated frequency sets, respectively, and $\Pi$ is the set of all permutations of length $L$.

RMSE quantifies frequency estimation accuracy but is meaningful only when $\hat{L}=L$ and a one-to-one correspondence between true and estimated components exists.
To complement this limitation, the normalized mean square error (NMSE) of signal reconstruction is employed as an alternative performance metric:
\begin{equation} \label{NMSE_general}
    \operatorname{NMSE}(\mathbf{x}) =
    \mathbb{E}\left[\frac{\|\mathbf{x}-\hat{\mathbf{x}}\|_2^2}{\|\mathbf{x}\|_2^2}\right],
\end{equation}
where $\hat{\mathbf{x}}$ denotes the reconstructed signal based on the estimated parameters according to \eqref{1D_channel} and \eqref{MD_channel}.

Unlike RMSE, NMSE remains well-defined when $\hat{L}\neq L$, thereby capturing errors in both parameter estimation and model order detection. 
Accordingly, NMSE is also used as the training loss for the proposed networks.

\section{Proposed NeuPLSE Framework} \label{NeuPLSE_framework}

\begin{figure*}[htbp]
    \centering
	\includegraphics[width=6.2in, trim=0 5pt 0 10pt, clip]{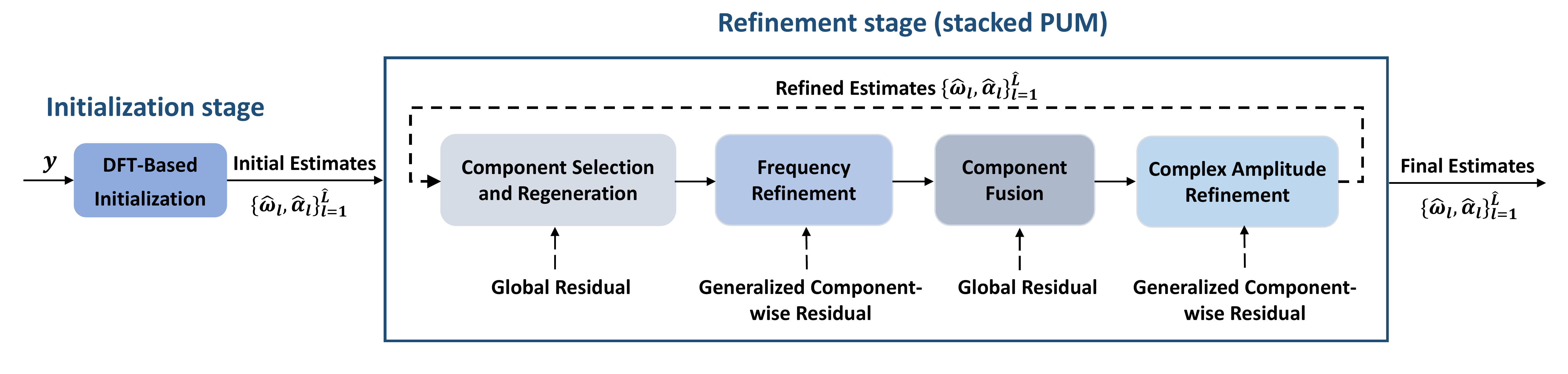}
    \vspace{-8pt}
    \caption{The unified framework of NeuPLSE.}
    \label{neuplse}
\end{figure*}

We introduce NeuPLSE, a unified framework for off-grid LSE applicable to both 1D and MD scenarios. 
As illustrated in Fig.~\ref{neuplse}, NeuPLSE adopts a two-stage architecture consisting of an initialization stage and a refinement stage, integrating model-driven signal structures with data-driven adaptability. 
The framework exhibits structural scalability, as it is developed under a unified architecture for both 1D and MD settings.

In the initialization stage, a DFT-based scheme is employed to obtain coarse estimates of spectral parameters and the model order. These initial estimates provide a starting point for the subsequent refinement stage, which consists of a sequence of Parameter Update Modules (PUMs) that jointly perform model order detection and parameter refinement based on residual analysis. Specifically, residual analysis is conducted at two complementary levels.

First, global residual analysis is performed to detect the model order. 
The global residual is defined as
\begin{equation}
\mathbf{r} = \mathbf{y} - \hat{\mathbf{x}},
\end{equation}
which represents the mismatch between the observation and its reconstruction.
Based on $\mathbf{r}$, model order detection is achieved through component selection, regeneration, and fusion. Specifically, selection removes components with negligible contribution, regeneration introduces new components based on residual information, and fusion merges redundant nearby components. These operations jointly enable adaptive model order detection while maintaining a parsimonious representation.

Second, generalized component-wise residual analysis is conducted for parameter refinement. 
A conventional component-wise residual is defined as
\begin{equation}
\mathbf{r}_l = \mathbf{y} - \sum_{s \neq l} \hat{\alpha}_s \mathbf{a}(\hat{\boldsymbol{\omega}}_s),
\end{equation}
which removes the contributions of all other components and provides a local observation for the $l$-th component. 
Instead of directly relying on this form, NeuPLSE constructs a generalized component-wise residual by adaptively combining the current component reconstruction, the global residual, and the contributions from other components. 
This generalized component-wise residual enables more flexible spectral parameter updates, with its explicit mathematical form presented in the following section.

NeuPLSE provides a unified framework for LSE across 1D and MD settings, forming the basis for the subsequent implementations.
For clarity, the main notation is summarized in Table~\ref{tab:nomenclature}, which serves as a reference for the following sections. 
For notational simplicity, the hat notation used in Fig.~\ref{neuplse} is omitted hereafter unless needed for clarity; the distinction between true and estimated quantities is made explicit when necessary.

\begin{table}[htbp]
    \centering
    \caption{Summary of Main Notation in NeuPLSE}
    \label{tab:nomenclature}
    \begin{tabular}{@{}l m{0.7\columnwidth}@{}}
        \toprule
        \textbf{Symbol} & \textbf{Description} \\
        \midrule
        
        \multicolumn{2}{@{}l}{\textit{Problem Setup and Signal Model}} \\
        \midrule
        $N, D$ & Number of observations and dimensionality. \\
        $L$ & Model order (number of spectral components). \\
        $\alpha_l$ & Complex amplitude of the $l$-th component. \\
        $\omega_l, \boldsymbol{\omega}_l$ & Frequency (scalar in 1D, vector in MD). \\
        $\mathbf{y}, \boldsymbol{\mathcal{Y}}$ & Observed signal. \\
        
        \midrule
        \multicolumn{2}{@{}l}{\textit{Network Structure and Indices}} \\
        \midrule
        $J, j$ & Number of PUMs and index. \\
        $I^{(j)}, i$ & Number of FAUS modules in the $j$-th PUM and index. \\
        $T, t$ & Number of DGS layers and index. \\
        $L^{(0)}$ & Model order after initialization. \\
        $L_1^{(j)}, L_2^{(j)}$ & Model orders after selection and regeneration, and after fusion, respectively. \\
        
        \midrule
        \multicolumn{2}{@{}l}{\textit{Intermediate Estimates and Residuals}} \\
        \midrule
        $\omega_l^{(0)}, \omega_l^{d,(0)}$ & Frequency estimates after initialization. \\
        $\alpha_l^{(0)}$ & Complex amplitude estimates after initialization. \\
        $\omega_l^{(j,i,t)}, \omega_l^{d,(j,i,t)}$ & Frequency estimates during refinement. \\
        $\alpha_l^{(j,i)}$ & Complex amplitude estimates during refinement. \\
        $\mathbf{\overline{y}}_{\mathrm{re}}^{(j)}, \boldsymbol{\overline{\mathcal{Y}}}_{\mathrm{re}}^{(j)}$ & Global residual. \\
        $\mathbf{\overline{y}}_{l}^{(j,i,t)}, \boldsymbol{\overline{\mathcal{Y}}}_{l}^{(j,i,t)}$ & Generalized component-wise residual. \\
        $\mathbf{P}_{\mathrm{re}}^{(j)}, \boldsymbol{\mathcal{P}}_{\mathrm{re}}^{(j)}$ & DFT spectrum of the global residual. \\
        
        \midrule
        \multicolumn{2}{@{}l}{\textit{Threshold Hyperparameters and Learnable Parameters}} \\
        \midrule
        $\beta_{\mathrm{sel}}^{\mathrm{ini}}, \beta_{\mathrm{sel}}^{\mathrm{ref}}$ & Selection thresholds in the initialization and refinement stages. \\
        $\beta_{\mathrm{fus}}^{\mathrm{ini}}, \beta_{\mathrm{fus}}^{\mathrm{ref}}$ & Fusion thresholds in the initialization and refinement stages for 1D. \\
        $\beta_{\mathrm{fus},d}^{\mathrm{ini}}, \beta_{\mathrm{fus},d}^{\mathrm{ref}}$ & Fusion thresholds along dimension $d$ in the initialization and refinement stages for MD. \\
        $\beta_{\mathrm{bir}}^{\mathrm{ref}}$ & Regeneration threshold in the refinement stage. \\
        $\lambda_{\omega,1}^{(j,i,t)}, \lambda_{\omega,2}^{(j,i,t)}$ & Learnable coefficients for frequency refinement. \\
        $\lambda_{\alpha,1}^{(j,i)}, \lambda_{\alpha,2}^{(j,i)}$ & Learnable coefficients for amplitude refinement. \\
        $\eta^{(j,i,t)}, \eta^{d,(j,i,t)}$ & Learnable step sizes for frequency updates. \\
        \bottomrule
    \end{tabular}
\end{table}

\section{1D Implementation of NeuPLSE} \label{Sec_1DNeuPLSE}

In this section, we present the 1D implementation of the NeuPLSE framework for 1D-LSE. The objective is to jointly estimate the model order $L$ and the corresponding frequency--amplitude pairs $\{(\omega_l,\alpha_l)\}_{l=1}^{L}$ from the observation $\mathbf{y}$. 
 
\subsection{Architecture of 1D-NeuPLSE} \label{1D-NeuPLSE Architecture}
 
\begin{figure*}[htbp]
	\centering
	\includegraphics[width=6.5in]{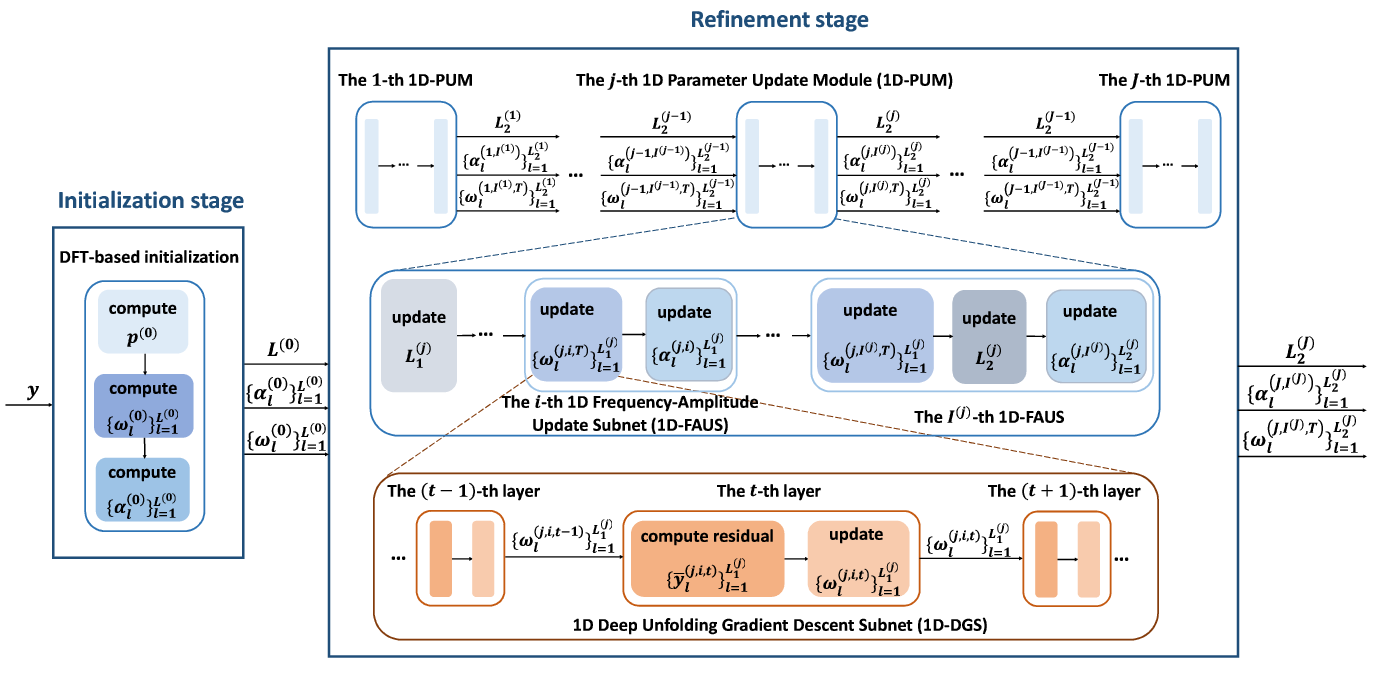}
    \vspace{-6pt}
	\caption{The architecture of 1D-NeuPLSE.}
	\label{Struct_1D_net}
\end{figure*}

Following the unified framework in Section~\ref{NeuPLSE_framework}, this section presents the 1D implementation of NeuPLSE. The overall architecture is illustrated in Fig.~\ref{Struct_1D_net}.

In the initialization stage, for \( \mathbf{y} \in \mathbb{C}^{N} \), a dictionary \( \mathbf{F}_\gamma \in \mathbb{C}^{\gamma N \times N} \) is formed from the first \( N \) columns of a \( \gamma N \times \gamma N \) DFT matrix. The DFT spectrum is then computed as
\begin{align} \label{initial_DFT}
\mathbf{p}^{(0)} = \mathbf{F}_\gamma \mathbf{y}.
\end{align}
Spectral candidates are identified by locating peaks in the normalized power spectrum $ \frac{\left[ |\mathbf{p}_{1}^{(0)}|^2, \dots, |\mathbf{p}_{\gamma N}^{(0)}|^2 \right]^T}{\max\limits_i |\mathbf{p}_i^{(0)}|^2}$ that exceed a predefined threshold $ \beta_{\mathrm{sel}}^{\mathrm{ini}}$, yielding the initial frequency estimates \( \{\tilde \omega_{l}^{(0)}\}_{l=1}^{\tilde L^{(0)}} \).
To mitigate spectral leakage, a fusion operation is applied to closely spaced frequency estimates. Specifically, if two frequency estimates $\tilde \omega_l^{(0)}$ and $\tilde \omega_m^{(0)}$ satisfies
\begin{equation} \label{frequency_distance_ini}
    \left| \tilde \omega_l^{(0)} - \tilde \omega_m^{(0)} \right| < \beta_{\mathrm{fus}}^{\mathrm{ini}},
\end{equation} 
they are merged into a single estimate given by $\frac{\tilde \omega_l^{(0)} + \tilde \omega_m^{(0)}}{2}$, with the resulting value wrapped to $[0, 2\pi)$. 
The fused estimates are denoted by $\{\omega_l^{(0)}\}_{l=1}^{L^{(0)}}$. Finally, the corresponding complex amplitudes $\{\alpha_{l}^{(0)}\}_{l=1}^{L^{(0)}}$ are estimated using element-wise linear minimum mean square error (LMMSE) as
\begin{align} \label{1D_initial_amplitude}
    \alpha_{l}^{(0)} = \sigma_{\alpha}^2 \mathbf{a}(\omega_{l}^{(0)})^H \left( \sigma_{\alpha}^2 \mathbf{a}(\omega_{l}^{(0)}) \mathbf{a}(\omega_{l}^{(0)})^H + \sigma^2 \mathbf{I} \right)^{-1} \mathbf{y},
\end{align}
where $\sigma_{\alpha}^2 = \mathbb{E} \left[ \frac{1}{L} \sum_{l=1}^{L} |\alpha_l|^2 \right]$ denotes the average energy of the complex amplitudes.

After initialization, the refinement stage applies $J$ stacked 1D-PUMs to iteratively refine the spectral estimates. 
For the $j$-th 1D-PUM, the input set $\{(\omega_l^{(j-1,I^{(j-1)},T)}, \alpha_l^{(j-1,I^{(j-1)})})\}_{l=1}^{L_2^{(j-1)}}$ is processed through component selection and regeneration to produce a refined candidate set. Components with sufficiently large normalized energy are retained according to
\begin{equation} \label{1D_selection_set}
    \mathcal{I}^{(j)} = \left\{ l \in \{1, \dots, L_2^{(j-1)}\} \;\middle|\; \frac{|\alpha_l^{(j-1,I^{(j-1)})}|^2}{\alpha_{\max}^{(j-1,I^{(j-1)})}} > \beta_{\mathrm{sel}}^{\mathrm{ref}} \right\},
\end{equation}
where $\alpha_{\max}^{(j-1,I^{(j-1)})} = \max_{1 \leq k \leq L_2^{(j-1)}} |\alpha_k^{(j-1,I^{(j-1)})}|^2$.
For $j > 1$, additional spectral components that were potentially missed are regenerated based on the global residual. Specifically, the global residual signal is computed as
\begin{equation} \label{1D_Component_regeneration_residual}
    \mathbf{\overline{y}}_{\mathrm{re}}^{(j)} = \mathbf{y} - \sum_{l \in \mathcal{I}^{(j)}} \alpha_l^{(j-1,I^{(j-1)})} \mathbf{a}(\omega_l^{(j-1,I^{(j-1)},T)}),
\end{equation}
and its DFT spectrum is obtained via
\begin{equation} \label{residual_spectrum}
    \mathbf{p}_{\mathrm{re}}^{(j)} = \mathbf{F}_\gamma \mathbf{\overline{y}}_{\mathrm{re}}^{(j)}.
\end{equation}
New candidate frequencies \( \{\omega_{\mathrm{re},l}^{(j)}\}_{l=1}^{L_{\mathrm{re}}^{(j)}} \) are then identified from the residual spectrum by thresholding with $\beta_{\mathrm{bir}}^{\mathrm{ref}}$, and the associated amplitudes \( \{\alpha_{\mathrm{re},l}^{(j)}\}_{l=1}^{L_{\mathrm{re}}^{(j)}} \) are obtained via element-wise LMMSE estimation.
The selected and regenerated components are then combined, yielding a total of $L_1^{(j)} = |\mathcal{I}^{(j)}| + L_{\mathrm{re}}^{(j)}$ candidates. The frequency--amplitude pairs are uniformly represented as
\begin{multline} \label{1D_merged_frequency_amplitude_pairs}
    \left\{ (\omega_l^{(j,0,0)}, \alpha_l^{(j,0)}) \right\}_{l=1}^{L_1^{(j)}} = 
    \left\{ \left( \omega_{\mathrm{re},l}^{(j)}, \alpha_{\mathrm{re},l}^{(j)} \right) \right\}_{l=1}^{L_{\mathrm{re}}^{(j)}} \cup \\
    \left\{ \left(\omega_l^{(j-1,I^{(j-1)},T)}, \alpha_l^{(j-1,I^{(j-1)})} \right) \right\}_{l \in \mathcal{I}^{(j)}} .
\end{multline}

\begin{figure}[htbp]
	\centering
	\includegraphics[width=2.9in, trim=0 0 0 5pt, clip]{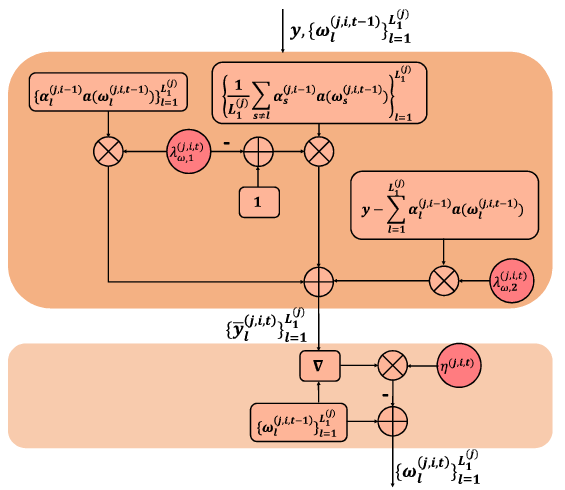}
    \vspace{-3pt}
	\caption{Computational flow of the $t$-th layer within the 1D-DGS for frequency refinement in the proposed 1D-NeuPLSE.}
	\label{omega_flow}
\end{figure}

The resulting candidate set is then refined by a cascade of 1D frequency-amplitude update subnet (FAUS) modules. Within the $i$-th 1D-FAUS of the $j$-th 1D-PUM, frequency refinement is performed via a $T$-layer 1D  deep unfolding gradient descent subnet (DGS). The computational flow for the $t$-th layer is illustrated in Fig. \ref{omega_flow}. Specifically, the generalized component-wise residual associated with the $l$-th component is computed as 
\begin{align} \label{1D_PUM_residual}
    {\mathbf{\overline{y}}}^{(j,i,t)}_{l} =  &\lambda_{\mathrm{\omega,1}}^{(j,i,t)} \alpha_l^{(j,i-1)} \mathbf a(\omega_l^{(j,i,t-1)}) + \notag \\ 
    & \lambda_{\mathrm{\omega,2}}^{(j,i,t)}(\mathbf y-\sum_{l=1}^{L_1^{(j)}} \alpha_l^{(j,i-1)} \mathbf a(\omega_l^{(j,i,t-1)}) + \notag \\ 
	& (1-\lambda_{\mathrm{\omega,1}}^{(j,i,t)})/ L_1^{(j)} (\sum_{s \neq l} \alpha_s^{(j,i-1)} \mathbf a(\omega_s^{(j,i,t-1)})),
\end{align}
where $\lambda_{\mathrm{\omega,1}}^{(j,i,t)}$ and $\lambda_{\mathrm{\omega,2}}^{(j,i,t)}$ are trainable parameters. 
This generalized component-wise residual contains three terms: the reconstructed signal of the target component, the global residual of the current reconstruction, and a scaled contribution from the remaining components. 
Together, these terms provide a localized residual representation for refining the $l$-th frequency, where $\lambda_{\mathrm{\omega,1}}^{(j,i,t)}$ and $\lambda_{\mathrm{\omega,2}}^{(j,i,t)}$ determine their relative contributions.
Using this generalized component-wise residual, the frequency estimate is refined through a learnable gradient descent step:
\begin{equation} \label{update_w_by}
	\omega_l^{(j,i,t)} =  \omega_l^{(j,i,t-1)} - \eta^{(j,i,t)} \frac{\partial f(\omega_l^{(j,i,t)}; {\mathbf{\overline{y}}}^{(j,i,t)}_{l})}{\partial \omega_l^{(j,i,t)}},
\end{equation}
where $\eta^{(j,i,t)}$ is a learnable step size that controls the update magnitude and influences the convergence behavior.
The objective function, which promotes spectral alignment, is defined as
\begin{equation}
    f(\omega_l^{(j,i,t)}; {\mathbf{\overline{y}}}^{(j,i,t)}_{l}) = \left| \mathbf a(\omega_l^{(j,i,t)})^H \mathbf{\overline{y}}^{(j,i,t)}_{l} \right|^{-2}.
\end{equation}
For simplicity, we introduce the shorthand
$\tilde{\omega} \triangleq \omega_l^{(j,i,t)}, \quad \tilde{\mathbf{y}} \triangleq \mathbf{\overline{y}}^{(j,i,t)}_{l}$,
so that the partial derivative of the objective with respect to $\tilde{\omega}$ is  
\begin{equation} \label{Gradient_equation}
	\frac{\partial f(\tilde{\omega}; \tilde{\mathbf{y}})}{\partial \tilde{\omega}} =
	\frac{-2 \left[ u(\tilde{\omega}; \tilde{\mathbf{y}}) \frac{\partial u(\tilde{\omega}; \tilde{\mathbf{y}})}{\partial \tilde{\omega}} + v(\tilde{\omega}; \tilde{\mathbf{y}}) \frac{\partial v(\tilde{\omega}; \tilde{\mathbf{y}})}{\partial \tilde{\omega}} \right]}{|\mathbf a(\tilde{\omega})^H \tilde{\mathbf{y}}|^3 \sqrt{u(\tilde{\omega}, \tilde{\mathbf{y}})^2 + v(\tilde{\omega}, \tilde{\mathbf{y}})^2}},
\end{equation}
where 
$u(\tilde{\omega}; \tilde{\mathbf{y}}) = \operatorname{Re}\!\left(\mathbf a(\tilde{\omega})^H \tilde{\mathbf{y}}\right)$, 
$v(\tilde{\omega}; \tilde{\mathbf{y}}) = \operatorname{Im}\!\left(\mathbf a(\tilde{\omega})^H \tilde{\mathbf{y}}\right) .$

\begin{figure}[htbp]
	\centering
	\includegraphics[width=2.9in, trim=0 0 0 8pt, clip]{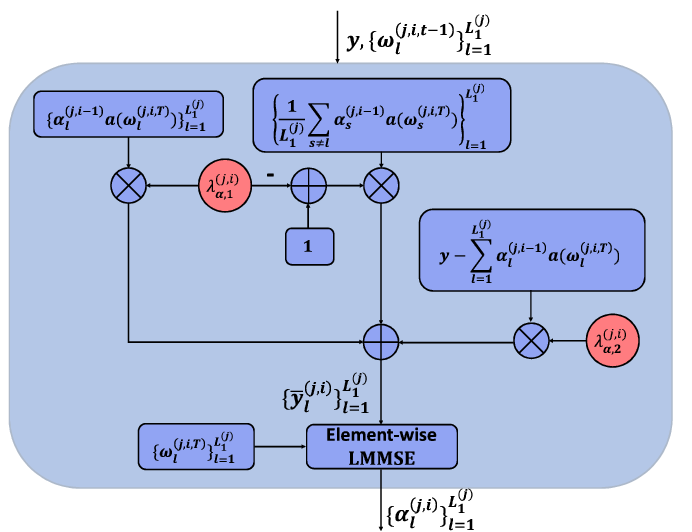}
	\caption{Computational flow of the amplitude update layer in the proposed 1D-NeuPLSE.}
	\label{alpha_flow}
\end{figure}

After frequency refinement, the complex amplitudes $\{\alpha_l^{(j,i)}\}_{l=1}^{L_1^{(j)}}$ are estimated via element-wise LMMSE, as illustrated in Fig.~\ref{alpha_flow}.  Specifically, the generalized component-wise residual of the $l$-th component is computed as
\begin{align} \label{1D_Amplitude_refinement}
	{\mathbf{\overline{y}}}^{(j,i)}_{l} = & \lambda_{\mathrm{\alpha,1}}^{(j,i)} \alpha_l^{(j,i-1)} \mathbf a(\omega_l^{(j,i,T)}) + \notag \\ 
    & \lambda_{\mathrm{\alpha,2}}^{(j,i)}(\mathbf y-\sum_{l=1}^{L_1^{(j)}} \alpha_l^{(j,i-1)} \mathbf a(\omega_l^{(j,i,T)}) + \notag \\ 
    & (1-\lambda_{\mathrm{\alpha,1}}^{(j,i)})/ L_1^{(j)} (\sum_{s \neq l} \alpha_s^{(j,i-1)} \mathbf a(\omega_s^{(j,i,T)})),
\end{align}
where $\lambda_{\mathrm{\alpha,1}}^{(j,i)}$ and $\lambda_{\mathrm{\alpha,2}}^{(j,i)}$ are learnable parameters.  
This generalized component-wise residual follows the same structural decomposition as in \eqref{1D_PUM_residual}, enabling amplitude update based on residual analysis, where $\lambda_{\mathrm{\alpha,1}}^{(j,i)}$ and $\lambda_{\mathrm{\alpha,2}}^{(j,i)}$ determine the relative contributions of different terms.
The complex amplitude is then obtained by
\begin{align} \label{1D_Amplitude_refinement_amplitude}
    \alpha_l^{(j,i)} = & \sigma_{\alpha}^2 \mathbf{a}(\omega_l^{(j,i,T)})^H \notag \\
	& \left( \sigma_{\alpha}^2 \mathbf{a}(\omega_l^{(j,i,T)}) \mathbf{a}(\omega_l^{(j,i,T)})^H + \sigma^2 \mathbf{I} \right)^{-1} \mathbf{\overline{y}}^{(j,i)}_{l}.
\end{align}

In the final $I^{(j)}$-th 1D-FAUS, a spectral fusion step is inserted before amplitude refinement to eliminate redundant components generated during iterative pursuit. Specifically, any two frequency estimates \( \omega_l^{(j,I^{(j)},T)} \) and \( \omega_m^{(j,I^{(j)},T)} \) whose separation satisfies
\begin{equation} \label{frequency_distance_ref}
  \left| \omega_l^{(j,I^{(j)},T)} - \omega_m^{(j,I^{(j)},T)} \right| < \beta_{\mathrm{fus}}^{\mathrm{ref}}
\end{equation}
are merged. The model order \( L_2^{(j)} \) is then updated by subtracting the number of merged pairs from the initial candidate count \( L_1^{(j)} \).

The threshold hyperparameters $\left[\beta_{\mathrm{sel}}^{\mathrm{ini}}, \beta_{\mathrm{fus}}^{\mathrm{ini}}, \beta_{\mathrm{sel}}^{\mathrm{ref}}, \beta_{\mathrm{fus}}^{\mathrm{ref}}, \beta_{\mathrm{bir}}^{\mathrm{ref}}\right]$ in 1D-NeuPLSE are specified based on empirical guidelines.
Specifically, $\beta_{\mathrm{sel}}^{\mathrm{ini}}$ is set to a normalized threshold of approximately $0.1$.
The initialization-stage fusion threshold $\beta_{\mathrm{fus}}^{\mathrm{ini}}$ is determined according to the spectral separation condition: 
in well-separated scenarios, it is set to the minimum inter-component spacing; in closely spaced scenarios, it is set to approximately twice the initial DFT grid spacing, i.e., $4\pi/(\gamma N)$, to facilitate component fusion.
The refinement-stage thresholds $\beta_{\mathrm{sel}}^{\mathrm{ref}}$ and $\beta_{\mathrm{fus}}^{\mathrm{ref}}$ follow a similar principle. 
In well-separated scenarios, $\beta_{\mathrm{sel}}^{\mathrm{ref}}$ is set to a normalized threshold of approximately $0.05$, and $\beta_{\mathrm{fus}}^{\mathrm{ref}}$ is set to the minimum inter-component spacing. 
In closely spaced scenarios, both thresholds are set to zero, thereby adaptively deactivating the corresponding mechanisms. 
Finally, the regeneration threshold $\beta_{\mathrm{bir}}^{\mathrm{ref}}$ is typically set within the range of $1$ to $4$, where a smaller value tends to yield lower residual energy by permitting more component regeneration.

\subsection{Overall Algorithm and Training Procedure} \label{Overall Algorithm and Training Procedure}

Algorithm~\ref{alg:1D_NeuPLSE_simple} summarizes the forward propagation of 1D-NeuPLSE. Starting from the initialization stage, the network proceeds through $J$ stacked 1D-PUMs for progressive refinement and outputs the final spectral estimates after the last PUM.

\begin{algorithm}
	\caption{Forward Propagation of the 1D-NeuPLSE }
	\label{alg:1D_NeuPLSE_simple}
	\begin{algorithmic}[1]
	\State \textbf{Input:} Received signal $\mathbf{y}$, oversampling factor $\gamma$, number of 1D-PUM $J$, number of 1D-FAUS $\{I^{(j)}\}_{j=1}^J$, number of layers $T$, thresholds $\left[\beta_{\mathrm{sel}}^{\mathrm{ini}}, \beta_{\mathrm{fus}}^{\mathrm{ini}}, \beta_{\mathrm{sel}}^{\mathrm{ref}}, \beta_{\mathrm{fus}}^{\mathrm{ref}}, \beta_{\mathrm{bir}}^{\mathrm{ref}}\right]$.

	\State Initialize $\{(\omega_{l}^{(0)},\alpha_{l}^{(0)})\}_{l=1}^{L^{(0)}}$ by \eqref{initial_DFT} to \eqref{1D_initial_amplitude}.

	\For{$j = 1$ to $J$}
		\State Update $\mathcal{I}^{(j)}$ by \eqref{1D_selection_set}.
		\If{$j > 1$}
            \State Update $\{(\omega_{\mathrm{re},l}^{(j)},\alpha_{\mathrm{re},l}^{(j)})\}_{l=1}^{L_{\mathrm{re}}^{(j)}}$ by \eqref{1D_Component_regeneration_residual} and \eqref{residual_spectrum}.
		\EndIf
		\State Update $\{(\omega_l^{(j,0,0)}, \alpha_l^{(j,0)})\}_{l=1}^{L_1^{(j)}}$ by \eqref{1D_merged_frequency_amplitude_pairs}.
		\For{$i = 1$ to $(I^{(j)}-1)$}
		    \For{$t = 1$ to $T$}
		         \State Update $\{\omega_l^{(j,i,t)}\}_{l=1}^{L_1^{(j)}}$ by \eqref{1D_PUM_residual} and \eqref{update_w_by}.
			\EndFor
			\State Update $\{\alpha_l^{(j,i)}\}_{l=1}^{L_1^{(j)}}$ by \eqref{1D_Amplitude_refinement} and \eqref{1D_Amplitude_refinement_amplitude}.
		\EndFor
		\For{$i = I^{(j)}$}
		    \For{$t = 1$ to $T$}
		         \State Update $\{\omega_l^{(j,I^{(j)},t)}\}_{l=1}^{L_1^{(j)}}$ by \eqref{1D_PUM_residual} and \eqref{update_w_by}.
			\EndFor
            \State Update $\{\omega_l^{(j,I^{(j)},T)}\}_{l=1}^{L_2^{(j)}}$ based on \eqref{frequency_distance_ref}.
			\State Update $\{\alpha_{l}^{(j,I^{(j)})}\}_{l=1}^{L_2^{(j)}}$ by \eqref{1D_Amplitude_refinement} and \eqref{1D_Amplitude_refinement_amplitude}.
		\EndFor
	\EndFor
	\State \textbf{Output:} $L_2^{(J)}$ and $\{(\omega_l^{(J,I^{(J)},T)},\alpha_l^{(J,I^{(J)})})\}_{l=1}^{L_2^{(J)}}$.
	\end{algorithmic}
\end{algorithm}

The 1D-NeuPLSE network is trained end-to-end by minimizing the NMSE loss introduced in Section~\ref{Problem Formulation}, i.e., the mismatch between the reconstructed signal $\hat{\mathbf{x}}_{\boldsymbol{\phi}}$ and the target signal $\mathbf{x}$:
\begin{align}
\mathcal{L}(\boldsymbol{\phi}) = \frac{  \left\| \hat{\mathbf{x}}_{\boldsymbol{\phi}} - \mathbf{x} \right\|^2 }{ \left\| \mathbf{x} \right\|^2 },
\end{align}
where $\boldsymbol{\phi} = \{\lambda_{\omega,1}^{(j,i,t)},\lambda_{\omega,2}^{(j,i,t)},\lambda_{\alpha,1}^{(j,i)},\lambda_{\alpha,2}^{(j,i)},\eta^{(j,i,t)}\}_{j,i,t}$ denotes all learnable parameters. These parameters are shared across all line spectral components, so that the network does not rely on a predefined maximum number of paths.    
The reconstructed signal is given by
\begin{align}
\hat{\mathbf{x}}_{\boldsymbol{\phi}} = \sum_{l=1}^{L_{2}^{(J)}}\alpha_{l}^{(J,I^{(J)})} \mathbf{a}(\omega_{l}^{(J,I^{(J)},T)}).
\end{align}
Training is performed using the Adam optimizer with a learning rate of $0.001$.

\subsection{Computational Complexity Analysis}

This subsection analyzes the computational complexity of 1D-NeuPLSE and compares it with that of 1D-VALSE, measured in terms of the number of complex multiplications.
The reported complexity covers the complete forward inference process of 1D-NeuPLSE.

The computational complexity of 1D-NeuPLSE can be decomposed into the initialization stage and the subsequent PUM-based refinement.
In the initialization stage, the number of complex multiplications is given by $\gamma N^2 + 0.25 \gamma N + L^{(0)}(2N+2) + \frac{5}{4}(\tilde L^{(0)}-L^{(0)})$.
For the component regeneration in the $j$-th 1D-PUM, the number of complex multiplications is $L_{\mathrm{re}}^{(j)}(\gamma N^2+(0.25\gamma+3.25)N+2) + 1.5N|\mathcal{I}^{(j)}|$. 
Additionally, for the $t$-th layer of the $i$-th 1D-FAUS in the $j$-th 1D-PUM, the number of complex multiplications for generalized component-wise residual construction and frequency refinement is $(14.75N+1)L_{1}^{(j)}$. 
Accordingly, the total computational complexity of 1D-NeuPLSE is given by $\gamma N^2 + 0.25 \gamma N + L^{(0)}(2N+2) + \frac{5}{4}(\tilde L^{(0)}-L^{(0)}) + \sum_{j=2}^{J}(L_{\mathrm{re}}^{(j)}(\gamma N^2+(0.25\gamma+3.25)N+2) + 1.5N|\mathcal{I}^{(j)}|) + \sum_{j=1}^{J}(\sum_{i=1}^{I^{(j)}}(T((14.75N+1)L_{1}^{(j)})+L_2^{(j)}(6.5N+2))
+\frac{5}{4}L_1^{(j)})$.

For comparison, the computational complexities of 1D-NeuPLSE and 1D-VALSE are summarized in Table~\ref{1D_Complexity}.
Since the detailed procedure of 1D-VALSE is not presented here, the following discussion focuses only on the dominant terms.
For both algorithms, the complexity consists of terms proportional to $N^2$, $N$, and constants, while the constant terms are negligible for large $N$.
For 1D-NeuPLSE, the quadratic term arises from the DFT-based initialization and the regeneration operations in subsequent 1D-PUMs, corresponding to the coefficient $\gamma\left(1+\sum_{j=2}^{J} L_{\mathrm{re}}^{(j)}\right)$.
In contrast, for 1D-VALSE, the quadratic term mainly originates from the sequential initialization over all candidate spectral components and the frequency updates performed across iterations, corresponding to the term $\frac{13}{4}N^2\!\left(L^{\mathrm{max}}+\sum_{j=1}^{J}L^{(j)}\right)$.
Accordingly, its $N^2$-term coefficient depends on both the number of initialized candidates $L^{\mathrm{max}}$ and the retained component numbers $\{L^{(j)}\}_{j=1}^{J}$ involved in iterative frequency refinement, which are typically much larger than those involved in 1D-NeuPLSE.
For the linear terms, the main part in 1D-NeuPLSE arises from a summation over the $J$ 1D-PUMs, with an inner summation over the number of 1D-FAUS modules $I^{(j)}$.
In 1D-VALSE, the main part of the linear-term complexity results from a summation over the $J$ iterations, with an inner summation over the retained components $L^{(j-1)}$ involved in each iteration for support-set updates and complex amplitude updates, yielding terms that depend on both $L^{(j-1)}$ and $L^{(j)}$.
Since these updates in 1D-VALSE are performed over the full retained component set in each iteration, whereas the corresponding inner summation in 1D-NeuPLSE is only over the 1D-FAUS modules within each 1D-PUM, the linear-term coefficient in 1D-VALSE is generally larger than that in 1D-NeuPLSE.
Moreover, 1D-NeuPLSE tends to require fewer iterations than 1D-VALSE, benefiting from the hybrid design.
Overall, 1D-NeuPLSE achieves substantially lower computational complexity than 1D-VALSE while maintaining iterative refinement based on residual analysis.

\begin{table}[htbp]
	\centering
	\caption{ The computational complexity }
	\label{1D_Complexity}
	\begin{tabular}{>{\centering\arraybackslash}m{1.5cm}|>{\centering\arraybackslash}m{6.5cm}}
		\toprule[1.1pt]  
		Algorithm & Complexity \\
		\hline 
		1D-NeuPLSE & $\mathcal{O}(\gamma N^2(\sum_{j=2}^{J}L_{\mathrm{re}}^{(j)}+1)+N(0.25\gamma+2L^{(0)}+\sum_{j=2}^{J}(0.25\gamma L_{\mathrm{re}}^{(j)}+\frac{3}{4}L_{\mathrm{re}}^{(j)}+1.5|\mathcal{I}^{(j)}|) + \sum_{j=1}^{J}(\sum_{i=1}^{I^{(j)}}(14.75TL_{1}^{(j)}+6.5L_{2}^{(j)}))+\sum_{j=1}^{J}(\sum_{i=1}^{I^{(j)}}(TL_{1}^{(j)}))$ \\
		\hline
		1D-VALSE & $ \mathcal{O}(\frac{13}{4}N^2(L^{\mathrm{max}} + \sum_{j=1}^{J} L^{(j)}) + N ( \frac{35}{2}L^{\mathrm{max}} + 2\sum_{l=1}^{L^{\mathrm{max}}} l + \sum_{j=1}^{J}  ( \sum_{l=1}^{L^{(j-1)}} (l^2 + 2l + \frac{5}{2} ) + \frac{43}{2}L^{(j)} + 2(L^{(j)})^2 ) ) + \sum_{l=1}^{L^{\mathrm{max}}} (3l^2 - \frac{3}{2}l + \frac{3}{2}) + \sum_{j=1}^{J} \sum_{l=1}^{L^{(j-1)}}(\frac{3}{2}l - l^3)) $ \\
		\bottomrule[1.1pt]
	\end{tabular}
\end{table}

\section{MD Implementation of NeuPLSE}  \label{Sec_MDNeuPLSE}

This section presents the MD implementation of NeuPLSE.
Because the core estimation mechanism is the same as in the 1D case, only the aspects that differ from the 1D implementation in Section~\ref{Sec_1DNeuPLSE} are discussed.

\subsection{Architecture of MD-NeuPLSE}

\begin{figure*}[htbp]
	\centering
	\includegraphics[width=6.5in]{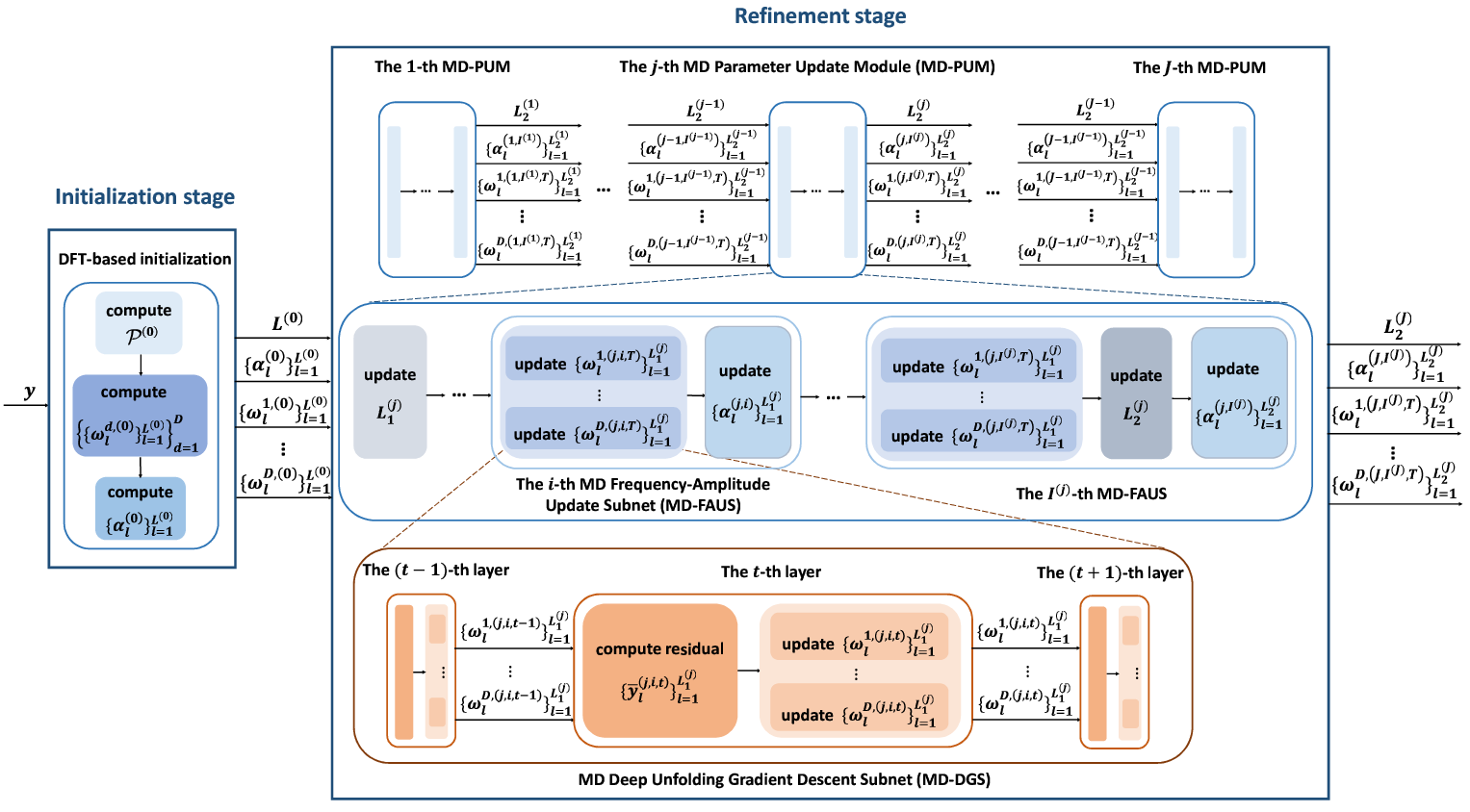}
	\caption{The architecture of MD-NeuPLSE.}
	\label{Struct_MD_net}
\end{figure*}

The proposed MD-NeuPLSE is developed under the unified framework presented in Section~\ref{NeuPLSE_framework} and aims to estimate the spectral parameters $\{\alpha_l,\omega_l^1,\dots,\omega_l^D\}_{l=1}^L$ without prior knowledge of $L$.
As illustrated in Fig.~\ref{Struct_MD_net}, the MD implementation also follows an initialization--refinement architecture.

In the initialization stage, the received signal $\mathbf{y}\in\mathbb{C}^{N}$ in \eqref{MD_measurements} is first reshaped into a $D$-dimensional tensor $\boldsymbol{\mathcal{Y}}\in\mathbb{C}^{N_1\times\dots\times N_D}$. For each dimension $d$, an oversampled DFT matrix $\mathbf{F}_{\gamma_d}\in\mathbb{C}^{\gamma_d N_d\times N_d}$ is constructed. The MD DFT spectrum is then computed through sequential mode-$d$ products
\begin{align} \label{MD_initial_DFT_tensor}
\boldsymbol{\mathcal{P}}^{(0)} = \boldsymbol{\mathcal{Y}} \times_1 \mathbf{F}_{\gamma_1} \times_2 \mathbf{F}_{\gamma_2} \cdots \times_D \mathbf{F}_{\gamma_D},
\end{align}
where $\times_d$ denotes the mode-$d$ product between a tensor and a matrix. 
Initial spectral candidates are obtained by detecting peaks in the normalized MD DFT spectrum whose power exceeds a threshold $\beta_{\mathrm{sel}}^{\mathrm{ini}}$, yielding frequency estimates $\{(\tilde\omega_l^1,\dots,\tilde\omega_l^D)\}_{l=1}^{\tilde L^{(0)}}$.
To mitigate spectral leakage, spectral fusion is applied independently along each dimension. For dimension $d$, if
\begin{equation} \label{frequency_distance_MD_ini}
    \left| \tilde \omega_l^{d,(0)} - \tilde \omega_m^{d,(0)} \right| < \beta_{\mathrm{fus},d}^{\mathrm{ini}},
\end{equation}
the two candidates are merged along that dimension.
After fusion across all dimensions, the model order becomes $L^{(0)}$ and the initial frequency set is $\{(\omega_l^{1,(0)},\dots,\omega_l^{D,(0)})\}_{l=1}^{L^{(0)}}$.
The corresponding amplitudes are estimated by MD-\eqref{1D_initial_amplitude}\footnote{``MD-(\(\cdot\))'' denotes the MD extension of the corresponding 1D expression, where scalar frequencies and steering vectors are replaced by their MD counterparts.}.

In the refinement stage, a sequence of $J$ MD-PUMs is applied to iteratively refine the spectral components. 
Given the input set $\{(\boldsymbol\omega_l^{(j-1,I^{(j-1)},T)}, \alpha_l^{(j-1,I^{(j-1)})})\}_{l=1}^{L_2^{(j-1)}}$, each MD-PUM first performs component selection according to the criterion in MD-\eqref{1D_selection_set} with threshold $\beta_{\mathrm{sel}}^{\mathrm{ref}}$. 
For $j>1$, additional components are then identified from the global residual defined in MD-\eqref{1D_Component_regeneration_residual}. 
The MD DFT spectrum of the global residual is computed as
\begin{align} \label{MD_initial_DFT_tensor_regeneration}
    \boldsymbol{\mathcal{P}}_{\mathrm{re}}^{(j)} = \overline{\boldsymbol{\mathcal{Y}}}_{\mathrm{re}}^{(j)} 
    \times_1 \mathbf{F}_{\gamma_1} \times_2 \mathbf{F}_{\gamma_2} \cdots \times_D \mathbf{F}_{\gamma_D},
\end{align}
where $\overline{\boldsymbol{\mathcal{Y}}}_{\mathrm{re}}^{(j)}$ denotes the tensor form of the residual signal.
The regenerated candidates $\{\boldsymbol\omega_{\mathrm{re},l}^{(j)}\}_{l=1}^{L_{\mathrm{re}}^{(j)}}$ are selected using threshold $\beta_{\mathrm{bir}}^{\mathrm{ref}}$, and their amplitudes are estimated via element-wise LMMSE.
These regenerated components are then merged with the selected components, yielding the candidate set $\{(\boldsymbol\omega_l^{(j,0,0)}, \alpha_l^{(j,0)})\}_{l=1}^{L_1^{(j)}}$.

The candidate set is further refined through a cascade of $I$ MD-FAUS modules. 
Within each MD-FAUS, frequency refinement is performed via a $T$-layer MD-DGS. 
At the $t$-th layer, the generalized component-wise residual associated with the $l$-th component is constructed following MD-\eqref{1D_PUM_residual}. 
The frequency parameters are updated independently along each dimension as
\begin{equation} \label{update_w_MD} 
	\omega_l^{d,(j,i,t)} = \omega_l^{d,(j,i,t-1)} - \eta^{d,(j,i,t)} \frac{\partial f(\boldsymbol \omega_l^{(j,i,t)}; \boldsymbol{\overline{\mathbf{y}}}^{(j,i,t)}_l)}{\partial \omega_l^{d,(j,i,t)}},
\end{equation}
where $\eta^{d,(j,i,t)}$ denotes the learnable step size for the $d$-th dimension, controlling the update magnitude along that dimension.
The objective function is defined as
\begin{equation}
f(\boldsymbol \omega_l^{(j,i,t)}; {\mathbf{\overline{y}}}^{(j,i,t)}_{l}) = \left| \mathbf a(\boldsymbol \omega_l^{(j,i,t)})^H \mathbf{\overline{y}}^{(j,i,t)}_{l} \right|^{-2}.
\end{equation}
After frequency refinement, the amplitudes $\{\alpha_l^{(j,i)}\}$ are updated by MD-\eqref{1D_Amplitude_refinement} and MD-\eqref{1D_Amplitude_refinement_amplitude}.

In the final $I^{(j)}$-th MD-FAUS, spectral fusion is performed before updating the complex amplitudes. 
For each dimension $d$, any two frequency estimates satisfying
\begin{equation} \label{frequency_distance_MD_ref}
    \left| \omega_l^{d,(j,I^{(j)},T)} - \omega_m^{d,(j,I^{(j)},T)} \right| < \beta_{\mathrm{fus},d}^{\mathrm{ref}}
\end{equation}
are merged in that dimension. 
This operation eliminates redundant components and yields the updated MD frequency set $\{(\omega_l^{1,(j,I^{(j)})}, \dots, \omega_l^{D,(j,I^{(j)})})\}_{l=1}^{L_2^{(j)}}$ with model order $L_2^{(j)}$.

\subsection{Overall Algorithm and Training Procedure}

The overall procedure of MD-NeuPLSE is summarized in Algorithm~\ref{alg:MD_NeuPLSE_simple}. 
Its learnable parameter set is
$\boldsymbol{\phi} = \left\{ \lambda_{\omega,1}^{(j,i,t)}, \lambda_{\omega,2}^{(j,i,t)}, \lambda_{\alpha,1}^{(j,i)}, \lambda_{\alpha,2}^{(j,i)}, \left\{ \eta^{d,(j,i,t)} \right\}_{d=1}^D \right\}_{j,i,t}$.
These parameters are shared across all line spectral components, so that no predefined maximum number of paths is required. 
The network is trained end-to-end using the NMSE loss defined in Section~\ref{Problem Formulation} and optimized by Adam with a learning rate of $0.001$.

\begin{algorithm}[htbp]
	\caption{Forward Propagation of the MD-NeuPLSE }
	\label{alg:MD_NeuPLSE_simple}
	\begin{algorithmic}[1]
	\State \textbf{Input:} Received signal $\mathbf{y}$, oversampling factors $\{\gamma_d\}_{d=1}^D$, number of MD-PUMs $J$, numbers of MD-FAUSs $\{I^{(j)}\}_{j=1}^J$, number of layers $T$, thresholds $\beta_{\mathrm{sel}}^{\mathrm{ini}}$, $\{\beta_{\mathrm{fus},d}^{\mathrm{ini}}\}_{d=1}^D$, $\beta_{\mathrm{sel}}^{\mathrm{ref}}$, $\{\beta_{\mathrm{fus},d}^{\mathrm{ref}}\}_{d=1}^D$, and $\beta_{\mathrm{bir}}^{\mathrm{ref}}$.

	\State Initialize $\{(\boldsymbol \omega_{l}^{(0)},\alpha_{l}^{(0)})\}_{l=1}^{L^{(0)}}$ by \eqref{MD_initial_DFT_tensor}, \eqref{frequency_distance_MD_ini}, and MD-\eqref{1D_initial_amplitude}.

	\For{$j = 1$ to $J$}
		\State Update $\mathcal{I}^{(j)}$ by MD-\eqref{1D_selection_set}.
		\If{$j > 1$}
            \State Update $\{(\boldsymbol \omega_{\mathrm{re},l}^{(j)},\alpha_{\mathrm{re},l}^{(j)})\}_{l=1}^{L_{\mathrm{re}}^{(j)}}$ by MD-\eqref{1D_Component_regeneration_residual} and \eqref{MD_initial_DFT_tensor_regeneration}.
		\EndIf
		\State Update $\{(\boldsymbol \omega_l^{(j,0,0)}, \alpha_l^{(j,0)})\}_{l=1}^{L_1^{(j)}}$ by MD-\eqref{1D_merged_frequency_amplitude_pairs}.
		\For{$i = 1$ to $(I^{(j)}-1)$}
		    \For{$t = 1$ to $T$}
		         \State Update $\{\boldsymbol \omega_l^{(j,i,t)}\}_{l=1}^{L_1^{(j)}}$ by MD-\eqref{1D_PUM_residual} and \eqref{update_w_MD}.
			\EndFor
			\State Update $\{\alpha_l^{(j,i)}\}_{l=1}^{L_1^{(j)}}$ by MD-\eqref{1D_Amplitude_refinement} and MD-\eqref{1D_Amplitude_refinement_amplitude}.
		\EndFor
		\For{$i = I^{(j)}$}
		    \For{$t = 1$ to $T$}
		         \State Update $\{\boldsymbol \omega_l^{(j,I^{(j)},t)}\}_{l=1}^{L_1^{(j)}}$ by MD-\eqref{1D_PUM_residual} and \eqref{update_w_MD}.
			\EndFor
            \State Update $\{\boldsymbol \omega_l^{(j,I^{(j)},T)}\}_{l=1}^{L_2^{(j)}}$ based on \eqref{frequency_distance_MD_ref}.
			\State Update $\{\alpha_{l}^{(j,I^{(j)})}\}_{l=1}^{L_2^{(j)}}$ by MD-\eqref{1D_Amplitude_refinement} and MD-\eqref{1D_Amplitude_refinement_amplitude}.
		\EndFor
	\EndFor
	\State \textbf{Output:} $L_2^{(J)}$ and $\{(\boldsymbol \omega_l^{(J,I^{(J)},T)},\alpha_l^{(J,I^{(J)})})\}_{l=1}^{L_2^{(J)}}$.
	\end{algorithmic}
\end{algorithm}

\subsection{Computational Complexity Analysis}

We now analyze the computational complexity of MD-NeuPLSE in terms of the number of complex multiplications over the complete forward inference process.
In the initialization stage, the computational cost is given by $\sum_{d=1}^{D} (\gamma_d (\prod_{i=1}^{d-1} \gamma_i) N_d N) + 0.25 \prod_{d=1}^{D} \gamma_d N + L^{(0)} (2 N + 2) + \frac{4+D}{4} (\tilde L^{(0)} - L^{(0)})$. 
For the $j$-th MD-PUM, the component regeneration requires $L_{\mathrm{re}}^{(j)} (\sum_{d=1}^{D} (\gamma_d (\prod_{i=1}^{d-1} \gamma_i) N_d N) + 0.25 \prod_{d=1}^{D} \gamma_d N + 3.25 N + 2) + (1.5 \sum_{d=1}^{D} N_d + \sum_{k=2}^{D} \prod_{d=1}^{k} N_d) |\mathcal{I}^{(j)}|$ complex multiplications. 
For each MD-FAUS layer, the generalized component-wise residual construction and frequency refinement incur $(\sum_{k=2}^{D} \prod_{d=1}^{k} 2 N_d + \sum_{d=1}^{D} N_d + N (7 + 2.75 D) + D) L_1^{(j)}$ complex multiplications.
Overall, the complexity of MD-NeuPLSE consists mainly of quadratic terms associated with DFT-based initialization and regeneration, and linear terms arising from the iterative refinement steps, while the constant terms are negligible. 
This complexity structure shows that the NeuPLSE framework remains computationally scalable in MD-LSE.

\section{Numerical Results} \label{sec:simulation}

In this section, we evaluate the performance of NeuPLSE for both its 1D and MD implementations in an ISAC application, adopted as a representative example without limiting the general applicability of the proposed framework. 
Comparisons with state-of-the-art baselines are first presented, followed by further analysis including loss curves, ablation studies, sensitivity evaluations, and generalization under mismatched distributions of path numbers.

Two typical settings are considered in the simulations: a sparse-path scenario with a small number of well-separated subpaths, and a dense-path scenario with many closely spaced subpaths.
Evaluation is carried out from the perspectives of estimation performance and computational efficiency. 
For the sparse-path scenario, estimation performance is assessed by the RMSE of the estimated parameters and the subpath number detection accuracy, whereas for the dense-path scenario, it is assessed by the NMSE of channel estimation. 
Computational efficiency is further evaluated in both scenarios in terms of the number of complex multiplications and runtime. 
The number of complex multiplications is adopted as the primary metric, since it provides a hardware-independent and consistent indicator of algorithmic cost; runtime is reported as a complementary measure of practical efficiency.
All methods are evaluated on the same CPU-based platform (Intel i9-14900K with 64 GB RAM).
For fair comparison, all baseline methods are carefully tuned for each scenario to achieve their best performance under the considered settings.

\subsection{1D-NeuPLSE}

In this subsection, we evaluate the performance of the proposed 1D-NeuPLSE.

\subsubsection{Sparse-path scenario}
 
The AoAs of the channel subpaths are independently drawn from a uniform distribution over $[-\pi,\pi]$, with a minimum separation of $15^\circ$ enforced to ensure resolvability. 
The complex gains are sampled from $\mathcal{CN}(1,0.1)$ with a minimum energy of $0.3$, and the number of subpaths $L$ is uniformly drawn from $\{1,\dots,5\}$. 
The base station (BS) employs a $128 \times 1$ uniform linear array (ULA) for AoA sensing. 
The considered baselines include classical model-based methods (CRLB \cite{8103120} , OMP~\cite{cai2011orthogonal}, VALSE~\cite{badiu2017variational}, and Superfast~\cite{hansen2018superfast}), data-driven approaches (Classifier-Estimator~\cite{naoumi2024complex} and Classifier-ComplexMLP~\cite{naoumi2024complex}), and deep unfolding methods (ADMM-Net~\cite{wang2024single} and ANM-Net~\cite{raza2025deep}).
For 1D-NeuPLSE, the structural hyperparameters are set to $J=2$, $I^{(1:2)}=\{3,3\}$, and $T=2$. The oversampling factor is set to $\gamma=2$, and the threshold parameters are specified as $\left[\beta_{\mathrm{sel}}^{\mathrm{ini}}, \beta_{\mathrm{fus}}^{\mathrm{ini}}, \beta_{\mathrm{sel}}^{\mathrm{ref}}, \beta_{\mathrm{fus}}^{\mathrm{ref}}, \beta_{\mathrm{bir}}^{\mathrm{ref}}\right]=[0.2,15^\circ,0.01,15^\circ,4]$.

\begin{figure}[htbp]
	\centering
	\includegraphics[width=2.9in]{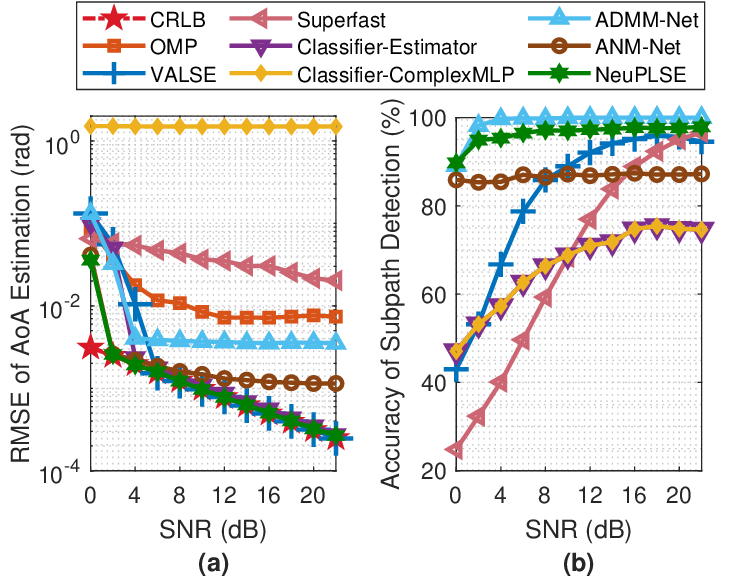}
	\captionsetup{skip=2pt}
	\caption{Estimation performance in the 1D sparse-path scenario. 
	(a) RMSE of AoA estimation versus SNR, conditioned on $\hat{L}=L$. 
	(b) Accuracy of subpath number detection versus SNR.}
	\label{fig:1D_sparse}
\end{figure}

Fig.~\ref{fig:1D_sparse} and Table~\ref{tab:complexity_table_1D_sparse} compare the estimation performance and computational efficiency of different methods.  
As shown in Fig.~\ref{fig:1D_sparse}(a), NeuPLSE and VALSE both achieve RMSE of the AoA estimation close to the CRLB at moderate-to-high SNRs, while NeuPLSE exhibits better robustness in the low-SNR regime.
Fig.~\ref{fig:1D_sparse}(b) further shows that NeuPLSE achieves subpath number detection accuracy above $90\%$ across the entire SNR range and outperforms most baselines. 
Although ADMM-Net attains slightly higher detection accuracy, its RMSE of AoA estimation is significantly worse, indicating inferior overall estimation performance. 
In terms of computational efficiency, NeuPLSE requires substantially fewer complex multiplications than several competitive baselines. 
In particular, the complex multiplication count of VALSE is 137 times that of NeuPLSE.
A similar trend is also observed in the runtime comparison.
Overall, these results demonstrate that NeuPLSE achieves a favorable tradeoff between estimation performance and computational efficiency.

\begin{table}[htbp] 
    \centering
    \caption{Comparison of computational efficiency in the 1D sparse-path scenario.}
    \label{tab:complexity_table_1D_sparse}
    \begin{tabular}{
        >{\centering\arraybackslash}p{2.76cm} |
        >{\centering\arraybackslash}p{2.81cm} |
        >{\centering\arraybackslash}p{1.85cm}
    }
    \Xhline{1.2pt}
    \textbf{Algorithm} & \textbf{Complex Mult. Count} & \textbf{Run Time (ms)} \\
    \Xhline{1.2pt}
	NeuPLSE & $1.19 \times 10^{5}$ & 5.15 \\
    \hline
    OMP & $1.00 \times 10^{5}$ & 2.58 \\
    \hline
    VALSE & $1.64 \times 10^{7}$ & 273.18 \\
    \hline
    Superfast & $3.97 \times 10^{5}$ &  7.13 \\
    \hline
    Classifier-Estimator & $5.53 \times 10^{6}$ & 100.55 \\
    \hline
    Classifier-ComplexMLP & $2.17 \times 10^{4}$ & 1.21 \\
    \hline
    ADMM-Net & $5.3 \times 10^{6}$ & 175.85 \\
    \hline
    ANM-Net & $2.7 \times 10^{5}$ &  10.90 \\
    \Xhline{1.2pt}
    \end{tabular}
\end{table}

\subsubsection{Dense-path scenario}

The channel is generated according to the CDL-B model \cite{3GPP_TS_38_901_v17.0.0} with two clusters, each consisting of 20 subpaths. Angular perturbations are introduced within each cluster to increase channel diversity.
The BS employs a $32 \times 1$ ULA for channel estimation.
The compared methods include classical model-based methods (OMP-LS~\cite{cai2011orthogonal}, VALSE~\cite{badiu2017variational}, Superfast~\cite{hansen2018superfast}, and Estimator-LS~\cite{naoumi2024complex}), pure data-driven approaches (UNet~\cite{marinberg2020study}, and Model2~\cite{melgar2022deep}), and deep unfolding methods (ADMM-Net~\cite{wang2024single} and ANM-Net~\cite{raza2025deep}). 
For 1D-NeuPLSE, the structural hyperparameters are configured as $J=2$, $I^{(1:2)}=\{4,4\}$, and $T=2$. The oversampling factor is set to $\gamma=4$, with threshold parameters $\left[\beta_{\mathrm{sel}}^{\mathrm{ini}}, \beta_{\mathrm{fus}}^{\mathrm{ini}}, \beta_{\mathrm{sel}}^{\mathrm{ref}}, \beta_{\mathrm{fus}}^{\mathrm{ref}}, \beta_{\mathrm{bir}}^{\mathrm{ref}}\right]=[0.05,7^\circ,0,0^\circ,1.5]$. 
In this dense scenario, the selection and fusion mechanisms of the refinement stage are adaptively deactivated, as indicated by $\beta_{\mathrm{sel}}^{\mathrm{ref}}=0$ and $\beta_{\mathrm{fus}}^{\mathrm{ref}}=0^\circ$.

\begin{figure}[htbp]
	\centering
	\includegraphics[width=3.1in, trim=0 0 0 5pt, clip]{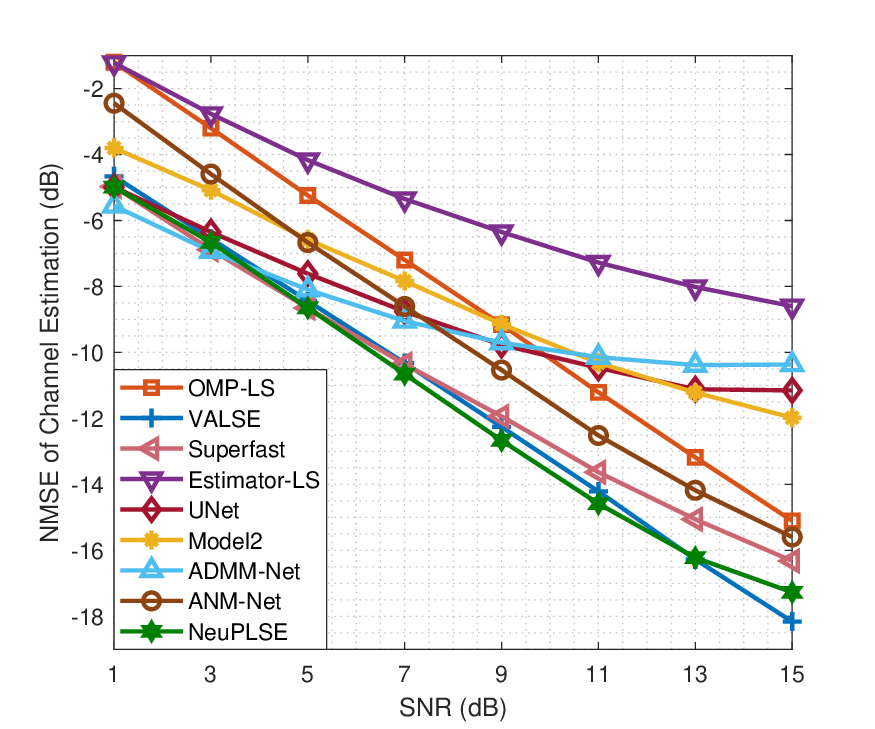}
	\captionsetup{skip=2pt}
	\caption{NMSE of channel estimation versus SNR in the 1D dense-path scenario.}
	\label{1D_CE}
\end{figure}

The channel estimation NMSE is shown in Fig.~\ref{1D_CE}, while the complex multiplication count and runtime are reported in Table~\ref{tab:complexity_table_1D_dense}. 
In the low-SNR regime, NeuPLSE maintains NMSE performance comparable to that of the best-performing baselines, whereas several other methods already exhibit clearly inferior results. 
As the SNR increases, VALSE remains the only baseline whose NMSE performance is comparable to that of NeuPLSE, while Superfast provides the most competitive performance among the remaining baselines, but still exhibits a clear performance gap relative to NeuPLSE. 
From the perspective of computational complexity, NeuPLSE requires fewer complex multiplications than most of the compared methods. 
Although UNet exhibits lower computational complexity, its NMSE performance is clearly inferior. 
In particular, VALSE requires nearly 50 times the number of complex multiplications required by NeuPLSE, while Superfast requires approximately 4 times the number required by NeuPLSE. 
A similar trend is also observed in the runtime results. 
These results show that NeuPLSE achieves NMSE performance comparable to that of VALSE with substantially reduced computational complexity and runtime, while maintaining superior estimation accuracy relative to the remaining baselines.

\begin{table}[htbp]
    \centering
    \caption{Comparison of computational efficiency in the 1D dense-path scenario.}
    \label{tab:complexity_table_1D_dense}
    \begin{tabular}{
        >{\centering\arraybackslash}p{2.76cm} |
        >{\centering\arraybackslash}p{2.81cm} |
        >{\centering\arraybackslash}p{1.85cm}
    }
    \Xhline{1.2pt}
    \textbf{Algorithm} & \textbf{Complex Mult. Count} & \textbf{Run Time (ms)} \\
    \Xhline{1.2pt}
	NeuPLSE & $7.06 \times 10^{4}$ & 3.43 \\
    \hline
    OMP-LS & $2.06 \times 10^{5}$ & 7.17 \\
    \hline
    VALSE & $3.52 \times 10^{6}$ & 115.29 \\
    \hline
    Superfast & $2.77 \times 10^{5}$ &  13.95 \\
    \hline
    Estimator-LS & $1.20 \times 10^{5}$ & 5.21 \\
    \hline
    UNet & $6.39 \times 10^{4}$ & 1.76 \\
    \hline
    Model2 & $2.01 \times 10^{5}$ & 5.50 \\
    \hline
    ADMM-Net & $6.67 \times 10^{5}$  &  47.36 \\
    \hline
    ANM-Net & $1.88 \times 10^{5}$ &  7.69 \\
    \Xhline{1.2pt}
    \end{tabular}
\end{table}

\subsection{MD-NeuPLSE}

In this subsection, we evaluate the performance of MD-NeuPLSE using its two-dimensional (2D) implementation as a representative case. 
For a fair comparison in the MD setting, baseline methods originally developed for 1D scenarios are extended in a consistent and straightforward manner by applying their original formulations along each dimension, without introducing additional modifications to their core algorithmic structures.

\subsubsection{Sparse-path scenario}

In this scenario, the AoAs and EoAs of all channel subpaths are generated independently from a uniform distribution over $[-\pi,\pi]$.
To guarantee angular resolvability, a minimum spacing of $15^\circ$ is imposed along each dimension.
The subpath number $L$ is randomly selected from $\{3,\dots,6\}$ with equal probability.
A $32\times16$ uniform rectangular array (URA) is deployed at the BS for joint AoA and EoA estimation.
The compared baselines include classical model-based methods (CRLB~\cite{8103120}, OMP~\cite{cai2011orthogonal}, and VALSE~\cite{badiu2017variational}), data-driven approaches (Classifier-Estimator~\cite{naoumi2024complex} and Classifier-ComplexMLP~\cite{naoumi2024complex}), and deep unfolding methods (ADMM-Net~\cite{wang2024single} and ANM-Net~\cite{raza2025deep}). 
For 2D-NeuPLSE, the structural parameters are set as $J=2$, $I^{(1:2)}=\{3,3\}$, and $T=2$. 
The oversampling factors are configured as $\gamma_1=2$, $\gamma_2=2$, and the threshold parameters are specified as $\{\beta_{\mathrm{sel}}^{\mathrm{ini}}, \beta_{\mathrm{fus},1}^{\mathrm{ini}}, \beta_{\mathrm{fus},2}^{\mathrm{ini}}, \beta_{\mathrm{sel}}^{\mathrm{ref}}, \beta_{\mathrm{fus},1}^{\mathrm{ref}}, \beta_{\mathrm{fus},2}^{\mathrm{ref}}, \beta_{\mathrm{bir}}^{\mathrm{ref}}\} = \{0.15,15^\circ,15^\circ,0.1,15^\circ,15^\circ,4\}$.

\begin{figure}[htbp]
	\centering
	\includegraphics[width=3.2in, trim=0 0 0 5pt, clip]{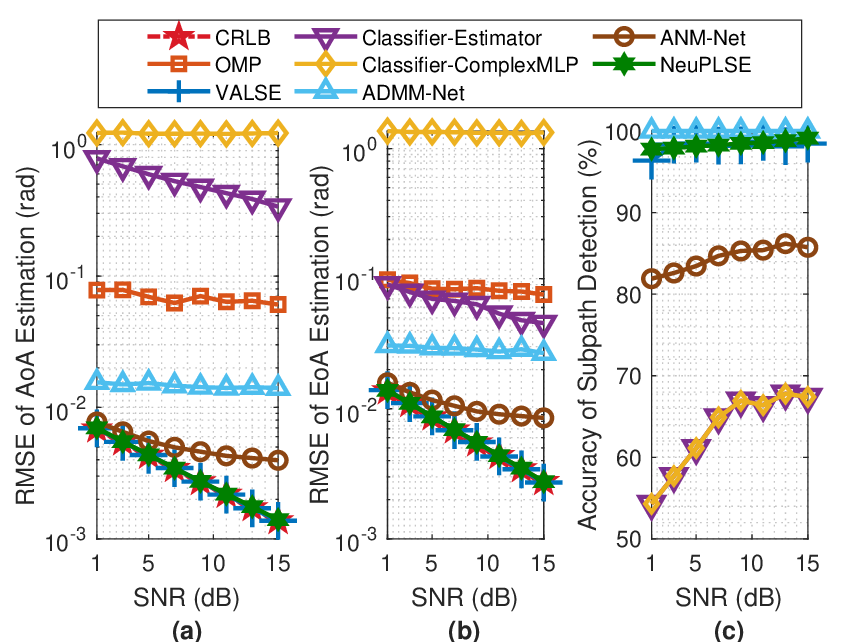}
	\captionsetup{skip=2pt}
	\caption{Estimation performance in the 2D sparse-path scenario. 
	(a) RMSE of AoA estimation versus SNR, conditioned on $\hat{L}=L$. 
	(b) RMSE of EoA estimation versus SNR, conditioned on $\hat{L}=L$. 
	(c) Accuracy of subpath number detection versus SNR.}
	\label{fig:2D_sparse}
\end{figure}

The estimation performance and computational efficiency comparison are shown in Fig.~\ref{fig:2D_sparse} and Table~\ref{tab:complexity_table_2D_sparse}, respectively. 
As shown in Fig.~\ref{fig:2D_sparse}(a) and (b), NeuPLSE and VALSE remain close to the CRLB for both AoA and EoA estimation RMSE over the entire SNR range, whereas the remaining methods exhibit larger errors. 
For subpath number detection in Fig.~\ref{fig:2D_sparse}(c), both NeuPLSE and VALSE maintain accuracy above $95\%$, with NeuPLSE exhibiting a slight advantage in the low-SNR regime. 
Although ADMM-Net attains higher detection accuracy, its AoA and EoA estimation errors remain considerably larger, leading to inferior overall estimation performance. 
In terms of complex multiplication count, NeuPLSE has lower computational complexity than most of the compared methods. 
Although Classifier-ComplexMLP has lower computational complexity, its estimation accuracy is clearly inferior. 
Notably, VALSE requires nearly one order of magnitude more complex multiplications than NeuPLSE. 
The runtime results show a consistent trend. 
Overall, NeuPLSE achieves high estimation accuracy and reliable subpath number detection with low computational complexity in the 2D sparse-path setting.

\begin{table}[htbp]
    \centering
    \caption{Comparison of computational efficiency in the 2D sparse-path scenario.}
    \label{tab:complexity_table_2D_sparse}
    \begin{tabular}{
        >{\centering\arraybackslash}p{2.76cm} |
        >{\centering\arraybackslash}p{2.81cm} |
        >{\centering\arraybackslash}p{1.85cm}
    }
    \Xhline{1.2pt}
    \textbf{Algorithm} & \textbf{Complex Mult. Count} & \textbf{Run Time (ms)} \\
    \Xhline{1.2pt}
	NeuPLSE & $6.37 \times 10^{5}$ & 7.64 \\
    \hline
    OMP & $5.28 \times 10^{6}$& 35.16 \\
    \hline
    VALSE & $5.58 \times 10^{6}$ & 71.98 \\
    \hline
    Classifier-Estimator & $5.08 \times 10^{7}$ & 317.08 \\
    \hline
    Classifier-ComplexMLP & $3.50 \times 10^{5}$ & 3.49 \\
    \hline
    ADMM-Net & $8.47 \times 10^{7}$ & 475.41 \\
    \hline
    ANM-Net & $9.21 \times 10^{6}$ &  55.94 \\
    \Xhline{1.2pt}
    \end{tabular}
\end{table}

\subsubsection{Dense-path scenario} 

The channel follows the CDL-B model \cite{3GPP_TS_38_901_v17.0.0} with three clusters, each comprising 20 subpaths. 
Intra-cluster angular variations are introduced to increase channel diversity.
At the user side, a $4 \times 1$ ULA is used for channel estimation over 198 subcarriers. 
The considered baselines include classical model-based methods (VALSE~\cite{badiu2017variational}, OMP-VALSE (a VALSE variant with OMP-based initialization), and Estimator-LS~\cite{naoumi2024complex}), pure data-driven approaches (UNet~\cite{marinberg2020study} and Model2~\cite{melgar2022deep}), as well as deep unfolding methods (ADMM-Net~\cite{wang2024single} and ANM-Net~\cite{raza2025deep}).
For 2D-NeuPLSE, the parameters are set as $J=4$, $I^{(1:4)}=\{5,4,3,3\}$, $T=2$, $\gamma_1=4$, and $\gamma_2=1$. The threshold parameters are configured as
$\left\{\beta_{\mathrm{sel}}^{\mathrm{ini}}, \beta_{\mathrm{fus},1}^{\mathrm{ini}}, \beta_{\mathrm{fus},2}^{\mathrm{ini}}, \beta_{\mathrm{sel}}^{\mathrm{ref}}, \beta_{\mathrm{fus},1}^{\mathrm{ref}}, \beta_{\mathrm{fus},2}^{\mathrm{ref}}, \beta_{\mathrm{bir}}^{\mathrm{ref}}\right\} = \{0.12,45^\circ,2.5^\circ,0,0^\circ,0^\circ,1.1\}$.  
In this dense setting, the refinement-stage selection and fusion mechanisms are adaptively deactivated, which is reflected by $\beta_{\mathrm{sel}}^{\mathrm{ref}}=0$ and $\beta_{\mathrm{fus},d}^{\mathrm{ref}}=0^\circ$ for $d=1,2$.

\begin{figure}[htbp]
	\centering
	\includegraphics[width=3.0in, trim=0 0 0 10pt, clip]{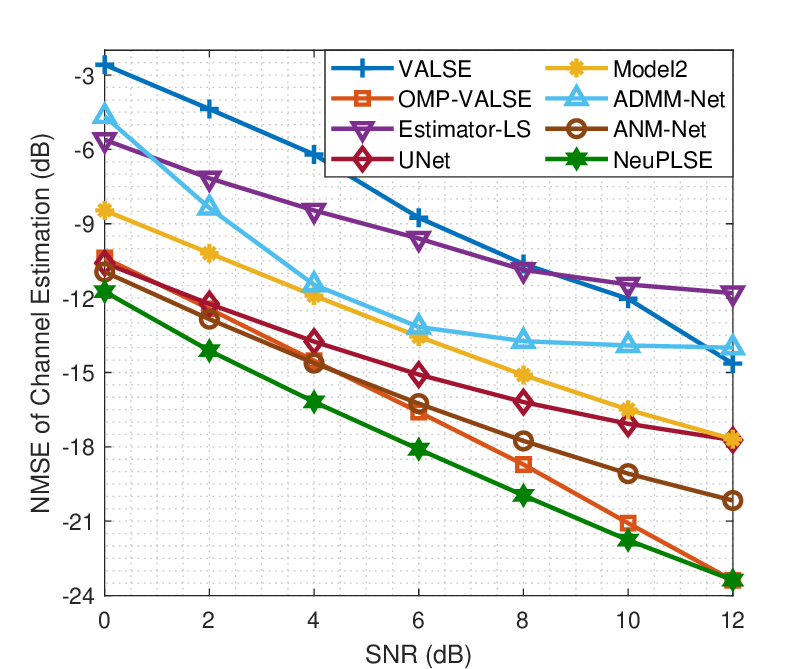}
	\captionsetup{skip=2pt}
	\caption{NMSE of channel estimation versus SNR in the 2D dense-path scenario.}
	\label{2D_CE}
\end{figure}

The NMSE results are shown in Fig.~\ref{2D_CE}, while the computational complexity and runtime are reported in Table~\ref{tab:complexity_table_2D_dense}.
NeuPLSE achieves the best NMSE performance over almost the entire SNR range. In particular, VALSE suffers from ineffective initialization in this dense scenario, which leads to clear performance degradation. Replacing its initialization with OMP improves the estimation accuracy and yields OMP-VALSE, which achieves NMSE performance comparable to that of NeuPLSE at $12$ dB but remains inferior in the low-to-moderate SNR regime.
The remaining baselines all exhibit worse NMSE performance than NeuPLSE, and their performance gaps generally become more pronounced as the SNR increases.
From the perspective of computational efficiency, NeuPLSE has the lowest computational complexity among all compared methods. In particular, OMP-VALSE requires approximately 45 times the number of complex multiplications required by NeuPLSE. The runtime results show a similar trend.
Overall, in the considered 2D dense-path scenario, NeuPLSE achieves both high channel estimation accuracy and low computational complexity.

\begin{table}[htbp]
    \centering
    \caption{Comparison of computational efficiency in the 2D dense-path scenario.}
    \label{tab:complexity_table_2D_dense}
    \begin{tabular}{
        >{\centering\arraybackslash}p{2.76cm} |
        >{\centering\arraybackslash}p{2.81cm} |
        >{\centering\arraybackslash}p{1.85cm}
    }
    \Xhline{1.2pt}
    \textbf{Algorithm} & \textbf{Complex Mult. Count} & \textbf{Run Time (ms)} \\
    \Xhline{1.2pt}
	NeuPLSE & $2.35 \times 10^{6}$ & 27.71 \\
    \hline
    VALSE & $3.42 \times 10^{9}$ & 13852.36 \\
    \hline
    OMP-VALSE & $1.08 \times 10^{8}$& 1441.45 \\
    \hline
    Estimator-LS & $4.22 \times 10^{8}$ & 810.39 \\
    \hline
    UNet & $7.15 \times 10^{6}$ & 8.94 \\
    \hline
    Model2 & $3.59 \times 10^{6}$ & 15.48 \\
    \hline
    ADMM-Net & $1.11 \times 10^{8}$ & 641.33 \\
    \hline
    ANM-Net & $8.67 \times 10^{6}$ & 41.23  \\
    \Xhline{1.2pt}
    \end{tabular}
\end{table}

The performance degradation observed at low SNRs in the 1D and MD evaluations is mainly associated with weaker component distinguishability in the DFT spectra and less reliable residuals during refinement. 
In the initialization stage, a higher noise level makes true spectral peaks less distinguishable from noise peaks, which may lead to missed or false components and less accurate coarse frequency estimates.
In the refinement stage, noise in the global residual reduces the reliability of component regeneration and model order update, while the residual of each component provides less accurate directions for frequency and amplitude refinement. 
This effect becomes more evident in dense scenarios, where closely spaced components cause stronger spectral overlap. 
In MD scenarios, errors along different frequency dimensions may jointly affect the final parameter estimates. 
Overall, the observed degradation is consistent with the increased ambiguity of the LSE problem under strong noise.

\subsection{Training and Validation Loss Curves}

To evaluate the convergence behavior and training stability of the proposed NeuPLSE, the training and validation loss curves are presented under a representative 2D dense scenario. 
The vertical axis denotes the loss value, defined as the NMSE of signal reconstruction, while the horizontal axis denotes the number of processed mini-batches.
The training data is generated using a mixed-SNR setting ranging from 0 to 12~dB, with 30,000 samples divided into training and validation sets at a ratio of 9:1.

\begin{figure}[htbp]
	\centering
	\includegraphics[width=3.2in, trim=0 0 0 5pt, clip]{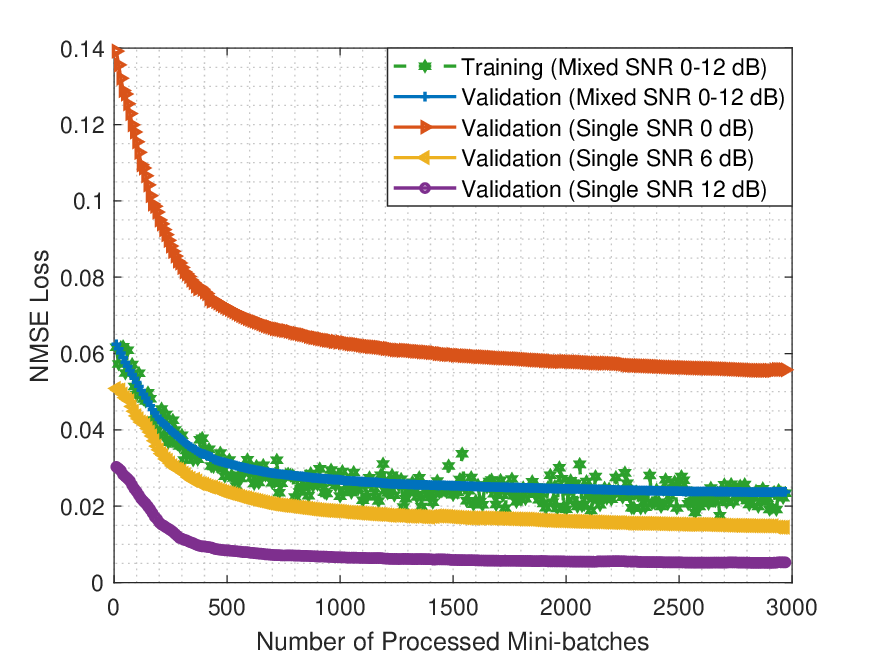}
	\captionsetup{skip=2pt}
	\caption{Training and validation loss curves under the 2D dense scenario with mixed SNR training.}
	\label{Loss_Curve}
\end{figure}

As shown in Fig.~\ref{Loss_Curve}, both training and validation losses exhibit stable convergence trends, with only mild fluctuations observed during training. 
These fluctuations are primarily attributed to the mixed-SNR training strategy, where mini-batches consist of samples with heterogeneous noise levels and varying estimation difficulty. 
Nevertheless, no instability or divergence is observed throughout the training process, indicating stable optimization behavior.
Moreover, the close alignment between training and validation losses indicates no evident overfitting.
To further assess robustness across different noise conditions, additional validation curves corresponding to 0~dB, 6~dB, and 12~dB are provided, all of which demonstrate consistent convergence behavior.
These results demonstrate that NeuPLSE achieves stable convergence under mixed-SNR training and maintains robustness across a wide range of SNR conditions.

NeuPLSE requires an offline training stage before deployment, with its cost governed by the number of processed mini-batches and the computation required within each mini-batch. 
For each mini-batch, the cost is mainly associated with the forward propagation of NeuPLSE and the update of a limited number of shared learnable algorithmic parameters under the NMSE objective. 
During online inference, these parameters remain fixed, and NeuPLSE performs only a single forward propagation per snapshot. 
Therefore, the real-time processing burden is characterized by the online per-snapshot inference complexity. 
When the signal statistics change, the trained model can be reused if the new operating conditions remain within the training distribution, while fine-tuning or retraining may be needed when the mismatch becomes substantial.

\subsection{Ablation Study of Core Structural Mechanisms}

To validate the necessity of the proposed structural mechanisms, ablation studies are conducted under a representative 2D sparse scenario. The ablation considers five mechanisms, namely initialization-stage selection and fusion, as well as refinement-stage selection, fusion, and regeneration. In this scenario, all five mechanisms are active, and each mechanism is individually removed to assess its contribution. In the full model, the adopted configuration is $I^{(1:2)}=\{3,3\}$ with $J=2$.
To further examine the role of the stacked FAUS structure, two variants are additionally considered:
i) fixing the number of PUMs with a single FAUS per PUM, i.e., $I^{(1:2)}=\{1,1\}$ with $J=2$;
ii) preserving the total number of FAUS layers by setting $I=1$ and increasing $J$, i.e., $I^{(1:6)}=\{1,1,1,1,1,1\}$ with $J=6$, thereby leading to more frequent structural updates while keeping the total number of parameter refinement steps unchanged.
For simplicity, the adopted configuration and the two stacked FAUS variants are denoted in the figure legends as $Adopted (I=3,J=2)$, $I=1, J=2$, and $I=1, J=6$, respectively.

\begin{figure}[htbp]
	\centering
	\includegraphics[width=3.2in]{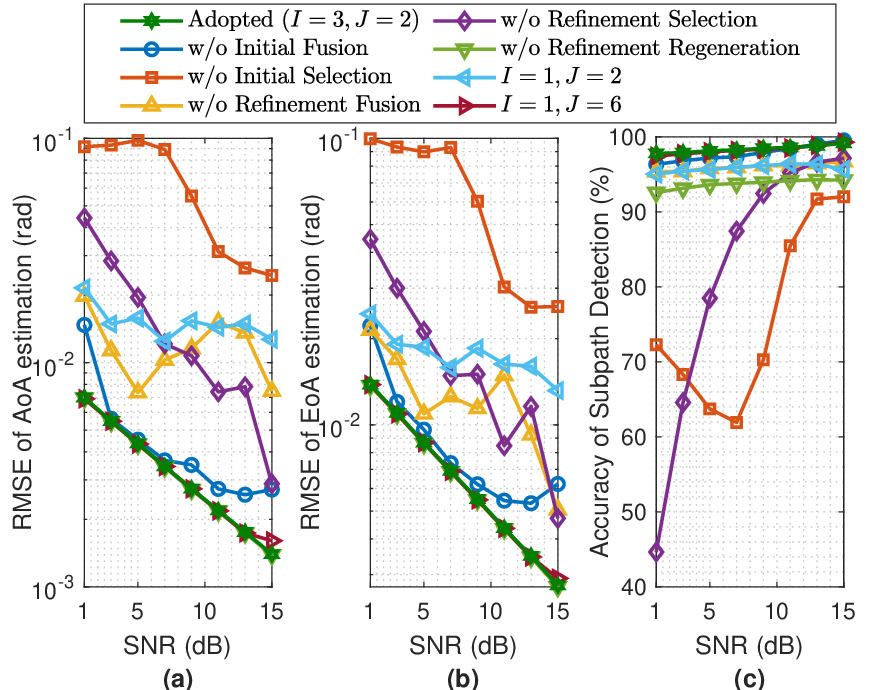}
	\captionsetup{skip=2pt}
	\caption{Ablation study of core structural mechanisms in the 2D sparse scenario.}
	\label{Ablation_Sparse}
\end{figure}

As shown in Fig.~\ref{Ablation_Sparse}, removing any of the five mechanisms degrades the performance in terms of both RMSE and path number detection accuracy, confirming the importance of all five mechanisms.
The variant with $I=1, J=6$ achieves performance close to that of the full model over most of the considered SNR range, but exhibits degraded AoA estimation accuracy at high SNRs and incurs higher computational cost due to more frequent structural updates.
The variant with $I=1, J=2$ shows consistently inferior results, indicating that insufficient stacked refinement limits the effectiveness of parameter updating.
The ablated variants also exhibit fluctuations across SNRs, indicating the role of these mechanisms in stabilizing estimation under different noise conditions; this is further supported by the threshold hyperparameter sensitivity analysis in Section~\ref{HSSA}, which shows that the proposed design remains robust over a reasonable range of threshold settings used to control these mechanisms.
Overall, these results demonstrate that all five mechanisms and the stacked FAUS design are essential to the proposed method.

\subsection{Sensitivity Analysis of Design Parameters} \label{HSSA}

To assess the robustness of the proposed NeuPLSE framework, sensitivity analyses are conducted on threshold hyperparameters, structural parameters, and DFT oversampling factors under a representative 2D sparse scenario.

\begin{figure}[htbp]
	\centering
	\includegraphics[width=3.3in]{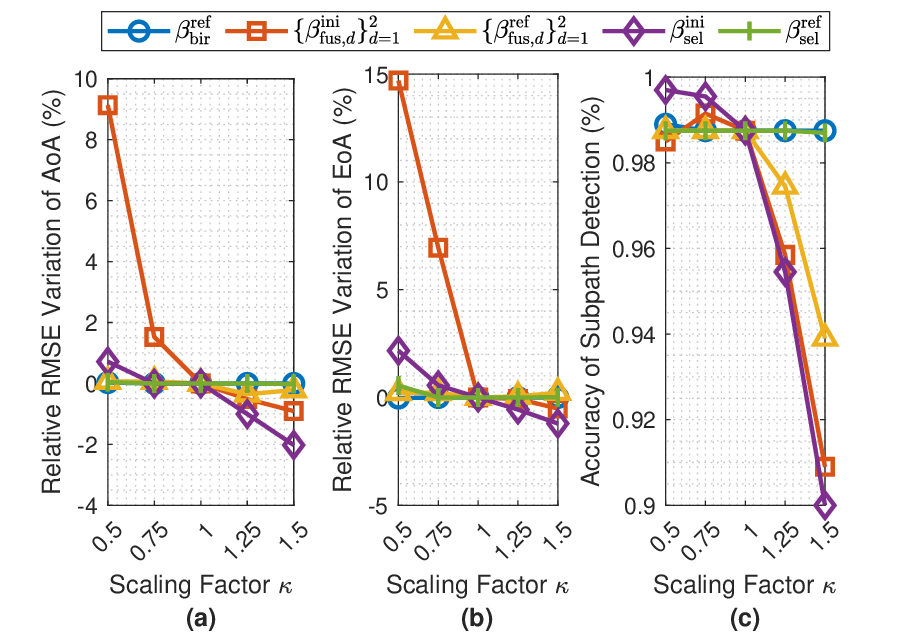}
    \caption{Sensitivity of threshold hyperparameters in the representative 2D sparse scenario.}
	\label{Sensitity_Beta_Sparse}
\end{figure}

The threshold hyperparameters govern the selection, fusion, and regeneration mechanisms, including
$\beta_{\mathrm{sel}}^{\mathrm{ini}}$, 
$\{\beta_{\mathrm{fus},d}^{\mathrm{ini}}\}_{d=1}^D$, 
$\beta_{\mathrm{sel}}^{\mathrm{ref}}$, 
$\{\beta_{\mathrm{fus},d}^{\mathrm{ref}}\}_{d=1}^D$, 
and $\beta_{\mathrm{bir}}^{\mathrm{ref}}$.
Their sensitivity is evaluated by scaling each threshold as $\theta=\kappa\theta^\star$, where $\theta^\star$ denotes the default value.
The fusion-related thresholds are scaled jointly across both dimensions.
As shown in Fig.~\ref{Sensitity_Beta_Sparse}, the vertical axis denotes the relative variation (in percentage) with respect to the default setting.
For $\kappa \in [0.5,1.5]$, the relative variations of AoA and EoA RMSE remain within $3\%$ for most threshold parameters, despite the large scaling range applied to the thresholds.
The initialization-stage fusion threshold $\beta_{\mathrm{fus}}^{\mathrm{ini}}$ exhibits relatively higher sensitivity, but its impact remains bounded within $15\%$.
The subpath number detection accuracy remains above $90\%$ across all settings.
Given that the AoA and EoA RMSE values under the default setting are on the order of $10^{-3}$, the observed relative variations correspond to only limited absolute changes.
These results indicate low sensitivity to the threshold hyperparameters and confirm the robustness of the associated selection, fusion, and regeneration mechanisms over a reasonably wide parameter range.

\begin{figure}[htbp]
	\centering
	\includegraphics[width=3.2in]{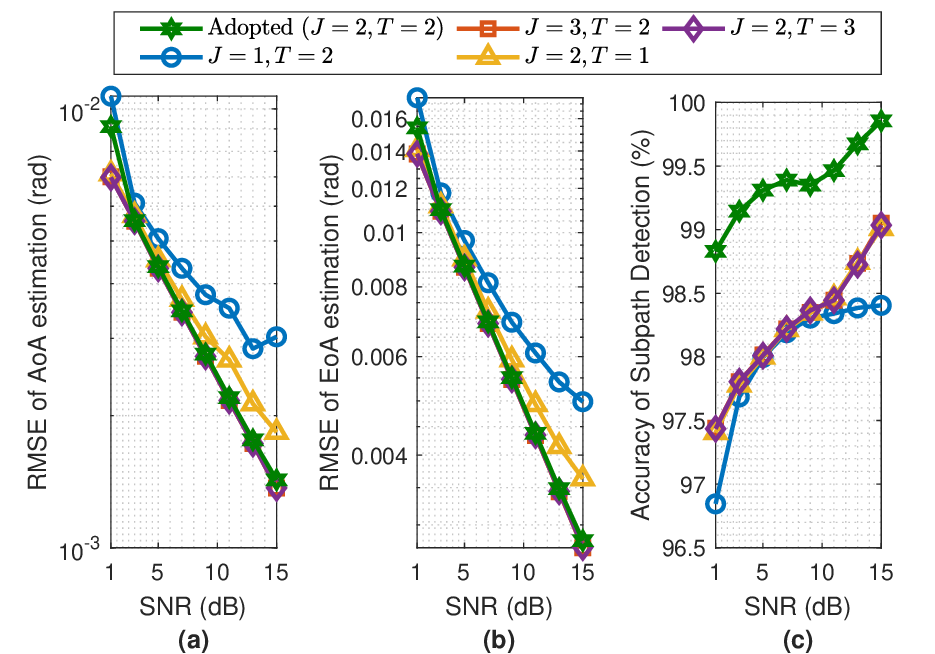}
	\captionsetup{skip=2pt}
	\caption{Sensitivity to Structural Parameters $(J,T)$ under the 2D sparse scenario.}
	\label{Sensitity_JT_Sparse}
\end{figure}

The structural parameters $(J, T)$ specify the number of PUMs and DGS layers, respectively, and thus determine the overall depth of the refinement process.
The sensitivity of the structural parameters is evaluated with respect to the adopted configuration $(J=2,T=2)$ by varying one parameter while fixing the other.
As shown in Fig.~\ref{Sensitity_JT_Sparse}, increasing $J$ or $T$ improves performance in the low-SNR regime, whereas reducing them leads to moderate degradation.
Nevertheless, the overall performance variation remains limited across the tested configurations.
The subpath number detection accuracy varies within $2\%$ and remains above $96.5\%$ across all tested configurations.
These results indicate low sensitivity to moderate changes in the structural parameters.

\begin{figure}[htbp]
	\centering
	\includegraphics[width=3.2in]{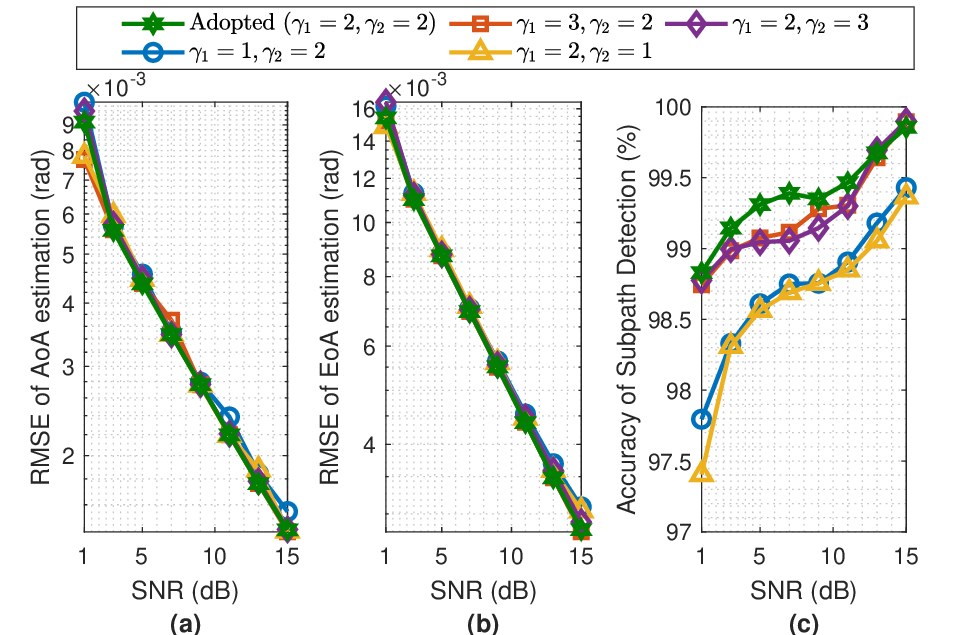}
    \caption{Sensitivity to the DFT oversampling factors in the 2D sparse scenario.}
	\label{Sensitity_Gamma_Sparse}
\end{figure}

The oversampling factors determine the initialization resolution.
Their sensitivity is evaluated by varying one factor while fixing the other, so as to isolate the effect of each factor on estimation performance.
In this scenario, the adopted oversampling factors are $\gamma_1 = 2$ and $\gamma_2 = 2$.
As shown in Fig.~\ref{Sensitity_Gamma_Sparse}, varying the oversampling factors causes only marginal changes in AoA and EoA RMSE, while the subpath number detection accuracy remains above $97\%$ across all settings.
This suggests that moderate changes in initialization resolution do not significantly affect the final performance.
Hence, the proposed framework exhibits low sensitivity to DFT oversampling factors.

\subsection{Generalization under Mismatched Path Number Distributions}

\begin{figure}[htbp]
	\centering
	\includegraphics[width=3.1in, trim=0 0 0 10pt, clip]{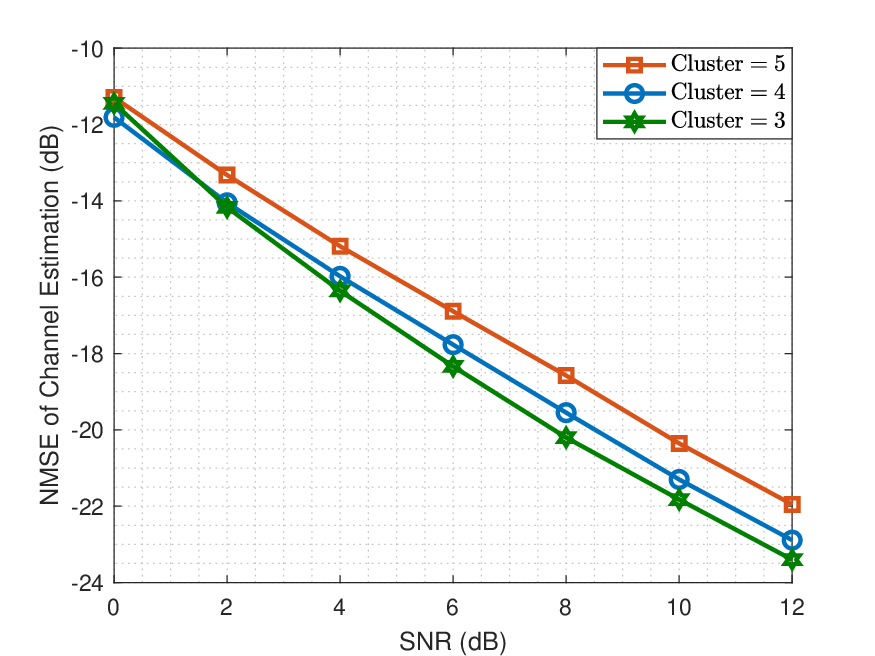}
	\captionsetup{skip=2pt}
	\caption{Generalization performance under increased cluster number in the 2D dense scenario.}
	\label{Generalization_PathNum_Dense}
\end{figure}

Generalization under mismatched path number distributions is further examined in a representative 2D dense scenario. 
The matched reference is obtained by using three clusters in the CDL-B model for both training and testing, with each cluster containing 20 subpaths. 
The mismatched cases use the same trained NeuPLSE model, while the number of clusters in testing is increased to four and five, corresponding to 80 and 100 propagation paths, respectively. 
As shown in Fig.~\ref{Generalization_PathNum_Dense}, the resulting NMSE degradation remains within approximately $1$~dB, indicating that NeuPLSE maintains stable channel estimation performance even under substantially denser propagation conditions. 
The observed degradation is associated with both the path number distribution mismatch and the increased estimation difficulty under denser propagation.
With more paths, the spectral density increases, and closely spaced components become less distinguishable in the DFT spectra. 
The residual after partial reconstruction also contains stronger contributions from missed components and neighboring components, which reduces the reliability of component regeneration and parameter refinement.

The generalization evaluation in Fig.~\ref{Generalization_PathNum_Dense} focuses on mismatch in the distribution of path numbers.
Other distribution shifts, such as changes in SNR, noise distribution, path separation, or frequency distribution, may influence the estimation performance by changing the distinguishability of spectral components and the reliability of residual information. 
Since NeuPLSE integrates model-driven signal structures with data-driven adaptability, moderate shifts within similar operating regimes may still be tolerable, although substantial mismatches may require further adaptation.

\section{Conclusion and Future Work} \label{sec:conclusion}

In this paper, we proposed NeuPLSE, a high-precision yet low-complexity framework for off-grid LSE under single-snapshot observations. 
By systematically integrating model-driven signal structures with data-driven adaptability, NeuPLSE enables joint model order detection and spectral parameter estimation through residual analysis. 
The proposed framework supports both 1D and MD scenarios through a unified design while maintaining low computational complexity. 
Extensive simulations in representative ISAC scenarios validated the effectiveness of NeuPLSE in terms of estimation accuracy and computational efficiency. 
Although the evaluation was conducted under ISAC scenarios, the proposed framework is generally applicable to a broad class of off-grid LSE problems.

While the current results are encouraging, the present study still relies on scenario-dependent training and hyperparameter configuration, which suggests the need for more automated configuration strategies and a more unified model with enhanced robustness across diverse scenarios in future work. 
In addition, evaluating NeuPLSE under more challenging lower SNR conditions, especially $\mathrm{SNR}<0$ dB, is also an important direction for further investigation. 
Moreover, extensions to nonideal conditions, including interference, array imperfections, synchronization errors, and dynamic channel variations, would provide a more comprehensive evaluation of the proposed framework.

\bibliographystyle{IEEEtran}
\bibliography{myRef}

\end{document}